\documentclass[sigconf,dvipsnames]{acmart}
\AtBeginDocument{%
  }

\copyrightyear{2025} 
\acmYear{2025} 
\setcopyright{cc}
\setcctype{by}
\acmConference[WPES '25]{Proceedings of the 2025 Workshop on Privacy in the Electronic Society}{October 13--17, 2025}{Taipei, Taiwan}
\acmBooktitle{Proceedings of the 2025 Workshop on Privacy in the Electronic Society (WPES '25), October 13--17, 2025, Taipei, Taiwan}\acmDOI{10.1145/3733802.3764058}
\acmISBN{979-8-4007-1898-4/2025/10}

\usepackage{dirtytalk}
\usepackage{multirow}
\usepackage{subcaption}

\begin{document}

\title[Understanding Contextual Vulnerability in Word-level Text Sanitization]{The Double-edged Sword of LLM-based Data Reconstruction: Understanding and Mitigating Contextual Vulnerability in Word-level Differential Privacy Text Sanitization}

\author{Stephen Meisenbacher}
\email{stephen.meisenbacher@tum.de}
\affiliation{%
  \institution{Technical University of Munich \\ School of Computation, Information and Technology}
  \city{Garching}
  \country{Germany}
}

\author{Alexandra Klymenko}
\email{alexandra.klymenko@tum.de}
\affiliation{%
  \institution{Technical University of Munich \\ School of Computation, Information and Technology}
  \city{Garching}
  \country{Germany}
}

\author{Andreea-Elena Bodea}
\email{andreea.bodea@tum.de}
\affiliation{%
  \institution{Technical University of Munich \\ School of Computation, Information and Technology}
  \city{Garching}
  \country{Germany}
}

\author{Florian Matthes}
\email{matthes@tum.de}
\affiliation{%
  \institution{Technical University of Munich \\ School of Computation, Information and Technology}
  \city{Garching}
  \country{Germany}
}

\renewcommand{\shortauthors}{Meisenbacher et al.}

\begin{abstract}
Differentially private text sanitization refers to the process of privatizing texts under the framework of Differential Privacy (DP), providing provable privacy guarantees while also empirically defending against adversaries seeking to harm privacy. Despite their simplicity, DP text sanitization methods operating at the \textit{word level} exhibit a number of shortcomings, among them the tendency to leave contextual clues from the original texts due to randomization during sanitization -- this we refer to as \textit{contextual vulnerability}. Given the powerful contextual understanding and inference capabilities of Large Language Models (LLMs), we explore to what extent LLMs can be leveraged to exploit the contextual vulnerability of DP-sanitized texts. We expand on previous work not only in the use of advanced LLMs, but also in testing a broader range of sanitization mechanisms at various privacy levels. Our experiments uncover a double-edged sword effect of LLM-based data reconstruction attacks on privacy and utility: while LLMs can indeed infer original semantics and sometimes degrade empirical privacy protections, they can also be used \textit{for good}, to improve the quality and privacy of DP-sanitized texts. Based on our findings, we propose recommendations for using LLM data reconstruction as a \textit{post-processing} step, serving to increase privacy protection by thinking adversarially.
\end{abstract}

\ccsdesc[500]{Security and privacy~Privacy protections}
\ccsdesc[500]{Security and privacy~Data anonymization and sanitization}
\ccsdesc[500]{Computing methodologies~Natural language processing}

\keywords{Differential Privacy, Natural Language Processing, LLM, Anonymization, Sanitization, Adversarial Inference, Data Reconstruction}


\maketitle

\vspace{-5pt}
\section{Introduction}
The application of Differential Privacy (DP) \cite{dwork2006differential} in Natural Language Processing (NLP) has been actively pursued by researchers in recent years, particularly to bring the robust guarantees promised by DP into the realm of language data processing, where the prevalence of sensitive data is particularly high. Incorporating DP guarantees directly on text data itself, referred to as \textit{differentially private text sanitization} \cite{yue-etal-2021-differential}, offers direct privacy guarantees on potentially sensitive texts, but the application of DP is not as straightforward as with more structured domains \cite{klymenko-etal-2022-differential}. Concretely, the discrete yet highly creative nature of language \cite{10.1145/3531146.3534642}, the high dimensionality required to represent language numerically \cite{feyisetan2021research}, and the procedures required to augment representations to adhere to the DP framework \cite{igamberdiev-habernal-2023-dp}, all lead to unique challenges in achieving DP NLP.

An early and intuitive class of solutions to fuse DP into NLP began at the base unit of language: the \textit{word} \cite{10.1145/3336191.3371856}. Word-level DP text sanitization approaches typically involve mapping an input word to a \say{noisy} output word that fulfills DP guarantees according to a chosen $\varepsilon$ parameter, known as the \textit{privacy budget}. This guarantee can be fulfilled in a number of ways, primarily either by adding noise to word embedding representations, or by using the Exponential Mechanism to randomize among a set of candidate replacement words \cite{hu-etal-2024-differentially}. Here, it is important to note that the privacy guarantee (i.e., adhering to DP) is given \textit{per word}, and that nearly all relevant literature leverages the notion of \textit{metric local} DP (MLDP), where sanitization is performed locally by the user \cite{meisenbacher2024comparative}.

A potential practical problem arises when considering the fact that word-level guarantees are not entirely useful in isolation, and that NLP tasks often necessitate full texts for implementing downstream tasks \cite{yue-etal-2021-differential}. Fortunately, DP allows for a transition from word-level to document-level guarantees, in that the property of \textit{compositionality} can be leveraged to reason about how word-level guarantees can be composed \cite{10.1145/3336191.3371856, 10.1145/3714393.3726504}. In particular, sanitizing all the $n$ tokens in a given document with privacy budget $\varepsilon = 3$ can be composed for a document-level budget of $3n$, leading to a unified, albeit weaker, guarantee from component sanitization steps. 

The major downside with respect to privacy when leveraging such compositionality was revealed in a recent work by \citet{tong-etal-2025-vulnerability}, who highlight the vulnerabilities introduced to word-level DP text sanitization when considering \textit{context}. The specific risks lie in the fact that while single word-level perturbations are performed in isolation, sanitizing (privatizing) an entire document leaves contextual clues which may aid a capable attacker in deducing both the overall original meaning, as well as in inferring original words based on context. This only exacerbates the challenges of word-level DP methods, which have already been criticized for their lack of contextual awareness \cite{mattern-etal-2022-limits}, wherein this limitation also opens the door to adversarial inference attacks.

Despite the important vulnerabilities demonstrated by \citet{tong-etal-2025-vulnerability}, we highlight three major limitations of the work. (1) Firstly, the authors only consider two mechanisms from a particular subclass of word-level DP mechanisms leveraging the Exponential Mechanism, namely \textsc{SanText} \cite{yue-etal-2021-differential} and \textsc{CusText} \cite{chen-etal-2023-customized}, thereby leaving it unknown how vulnerable other DP methods may be. (2) Secondly, according to our understanding of the work, in the execution of the sanitization steps, the authors do not set uniform document-level budgets, but instead choose word-level $\varepsilon$ values that are applied in an unbounded fashion to all texts (more on this in Section \ref{sec:case}), thereby not ensuring a fair comparison across all texts in a dataset. (3) Finally, the authors rely on BERT-based inference attacks, not exploring the inferential abilities of more advanced generative LLMs. 

We address these key limitations, with the goal of extending the findings of \citet{tong-etal-2025-vulnerability} and broadening the understanding and extent of \textit{contextual vulnerability} in word-level DP text sanitization methods. Following an initial reproduction case study with \textsc{SanText}, we implement LLM-based adversarial inference attacks on three additional word-level DP methods across a variety of $\varepsilon$ budgets, with the ultimate goal of (original) data reconstruction. We test the vulnerability of all four methods on three datasets, measuring privacy vulnerability using semantic and task-based metrics. Based on these results, we conduct a critical analysis, leading to recommendations for the field going forward.

Our findings reveal that although LLM-based data reconstruction attacks on DP-sanitized texts can serve to recapture semantics, degrade utility, and sometimes increase attacker performance, doing so also can prove to be beneficial as a \textit{post-processing} step after sanitization. In particular, we find that in some settings, using LLMs after DP text sanitization \textit{strengthens defense} against adversarial attacks, \textit{increases plausible deniability}, and \textit{improves the privacy-utility trade-off}, while always significantly improving text coherence. These results ground an analysis of the dangers and merits of DP text sanitization, leading to a set of practical recommendations.

We advance DP text privatization by widening the understanding of the \textit{contextual vulnerability} of word-level MLDP. Specifically:

\begin{enumerate}
    \vspace{-2pt}
    \itemsep 0em
    \item We advance recent literature on privacy vulnerabilities in word-level DP text sanitization by introducing a simple yet intuitive attack leveraging modern LLMs.
    \item We extend the testing of such vulnerabilities by conducting experiments with further methods, datasets, and metrics.
    \item We interpret our experimental results in a discussion on the dangers of word-level text sanitization methods for document privatization, but also include recommendations for their safe and effective use.
\end{enumerate}

\section{Foundations}
\subsection{Differential Privacy}
Differential Privacy \cite{dwork2006differential} is a mathematically grounded notion of privacy, specifically one that protects individuals by bounding the probability of inferring the information attributed to them in a dataset. Intuitively, such protections are offered by adding calibrated noise to computations on the data, or even the data itself, thereby granting plausibility deniability as to what the \say{true} value of the data may be. DP was originally envisioned for structured, relational datasets, in which each entry in the dataset represents the \textit{individual}, and \textit{neighboring} or \textit{adjacent} datasets can be defined as two datasets differing in exactly one individual data point (row). Under DP, any computations performed on two neighboring datasets should be \textit{indistinguishable} to some bound, which is governed by the privacy parameter $\varepsilon$, also known as the \textit{privacy budget}. Formally, this indistinguishability requirement is represented in the fundamental inequality enforced by DP: 
\[
    \frac{Pr[\mathcal{M}(D_1) \in \mathcal{S}]}{Pr[\mathcal{M}(D_2) \in \mathcal{S}]}
    \leq e^{\varepsilon}, 
\]
\noindent for any databases $D_1$ and $D_2$ differing in exactly one element, any $\varepsilon > 0$, any computation or function $\mathcal{M}$, and all $\mathcal{S} \subseteq Range(\mathcal{M})$.

This is known as $\varepsilon$-DP, and the notion above refers to \textit{global} DP. Another notion is that of \textit{local} DP (LDP) \cite{4690986}. In the local setting, we assume that the central curator or processor of the dataset is not trusted. Here, DP can be ensured at the user level; however, since the entirety of the dataset is not yet known, LDP imposes a much stricter indistinguishability requirement, i.e., between \textit{any} potential neighbor. This differs from the global notion, since neighboring databases only refer to those resulting from the dataset $D$. Formally, for the finite spaces $\mathcal{P}$ and $\mathcal{V}$, and for all $x, x' \in \mathcal{P}$ and all $z \in \mathcal{V}$:
\[
\frac{Pr[\mathcal{M}(x) = z]}{Pr[\mathcal{M}(x') = z]} \le e^\varepsilon
\]
Thus, an observed output cannot be attributed to a specific input with a high probability. While this notion is clearly stricter, it allows for the quantification of a privacy guarantee on the local, single data point level without the need for an aggregated dataset.

Another development in the DP field came with the growing need for reasoning about privacy in non-structured data settings, such as location privacy or in natural language. Introduced by \citet{chatzikokolakis2013broadening}, the notion of \textit{metric} DP augments the original definition by adapting the privacy reasoning to \textit{metric spaces}, or representational spaces endowed with a distance metric $d$. As can be seen in the definition of metric \textit{local} DP (MLDP), proposed by \citet{alvim2018local}, the indistinguishability requirement between any two data points is now scaled based on their distance (often interpreted as \textit{similarity}), thereby relaxing the strict, uniform requirement between \textit{any} two points as enforced by local DP:

\begin{displaymath}
   \frac{Pr[\mathcal{M}(x) = z]}{Pr[\mathcal{M}(x') = z]} \le e^{\varepsilon d(x,x')}
\end{displaymath}

In this work, we conduct experiments solely on mechanisms leveraging MLDP, as this has become the predominant notion for word-level DP, explained in the following.

\subsection{(Word-level) DP in NLP}
Early works conducting research on the integration of DP into NLP acknowledged several challenges \cite{feyisetan2021research, klymenko-etal-2022-differential}, chief of which was the transfer of DP concepts such as the \textit{individual} to the unstructured domain of text and language. With the aid of MLDP, many early approaches worked on differentially private text privatization on the \textit{word} level, leveraging the popular technique of word embeddings \cite{mikolov2013efficientestimationwordrepresentations} that were well suited to MDP notions.

Using different methods of word embedding representation, early word-level MLDP approaches designed \textit{perturbation} mechanisms, in which calibrated noise is added to word embedding representations, and then a nearest neighbor search is performed to reach a discrete \say{noisy} word \cite{fernandes2019generalised,10.1145/3336191.3371856}. Contemporary and follow-up works focused on optimizing the privacy-utility trade-off via the use of alternative metric spaces \cite{feyisetan3}, different distance metrics beyond the typical Euclidean distance \cite{xu2020differentially}, or modified mechanism designs \cite{xu3,carvalho2023tem,10.1145/3643651.3659896}. 

In another class of word-level MLDP approaches, other works leverage the fundamental Exponential Mechanism \cite{4389483} for DP word perturbations, extending MLDP in a \textit{utility-optimized} manner (UMLDP) \cite{yue-etal-2021-differential,chen-etal-2023-customized}. Specifically, such works ground their mechanisms in the notion that not all words are semantically sensible replacements for a given word, and thus, the replacement sets should be constrained to a smaller subset. Using the Exponential Mechanism, a noisy replacement can be chosen, scaled by the semantic similarity between each replacement candidate and the target word.

\begin{figure}[t!]
    \centering
    \includegraphics[scale=0.38]{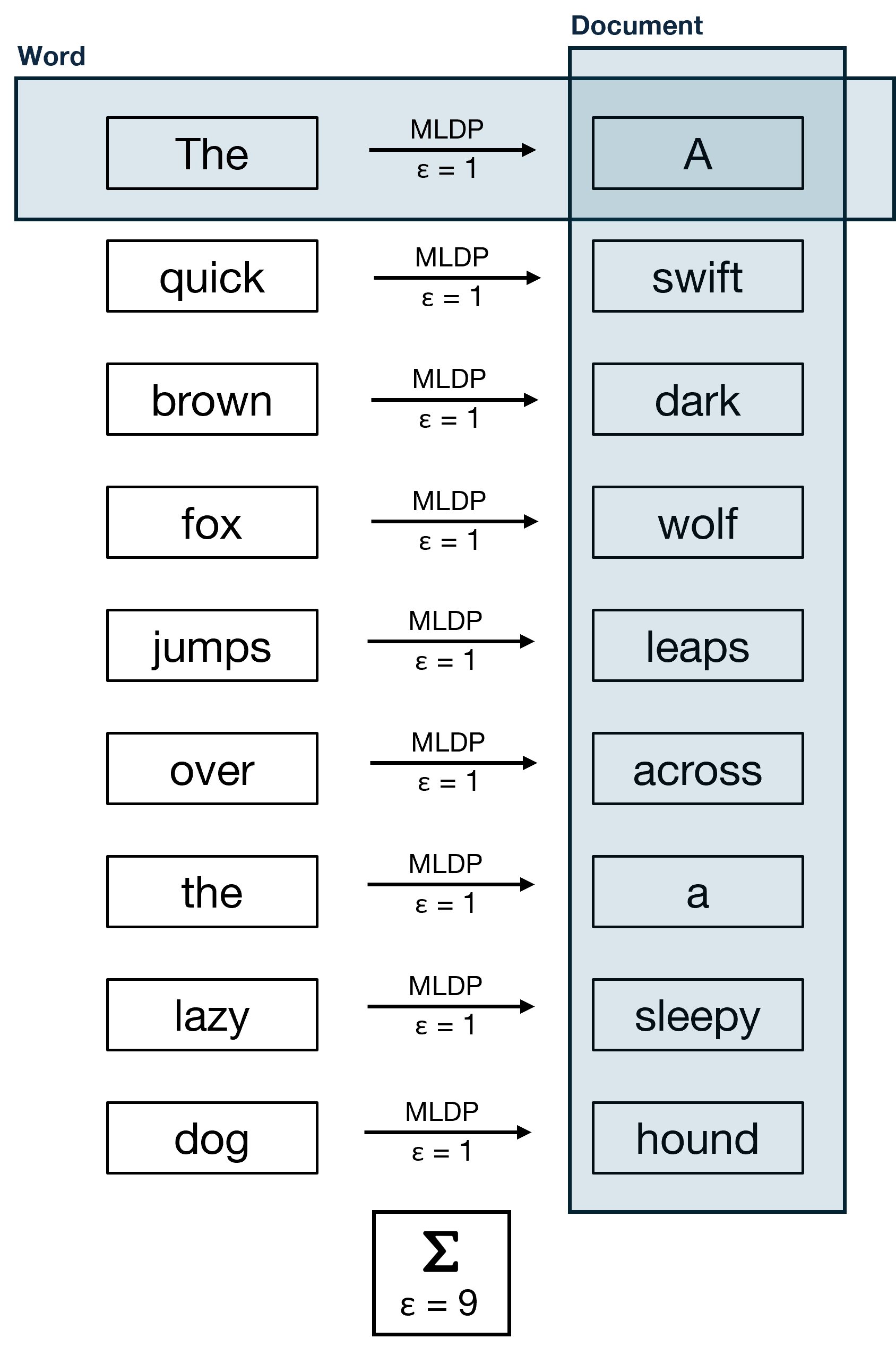}
    \caption{An example of word-level MLDP and document-level composition. Word-level MLDP (and the resulting privacy guarantees) operate \textit{per word}. In order to sanitize \textit{documents}, the basic composition property of DP is leveraged to achieve a document-level privacy budget and guarantee.}
    \label{fig:w2d}
\end{figure}

\subsubsection{From words to documents}
\label{sec:w2d}
An immediate limitation of the word-level model is the reasoning about the word as the \say{database}. Local DP necessitates that any two words are adjacent, and MLDP mechanisms thus only operate on (and provide privacy guarantees for) single words. This \textit{one user, one word} model \cite{10.1145/3336191.3371856} is quite limiting, as the vast majority of NLP tasks require larger units of data for meaningful analysis. Fortunately, the \textit{compositionality} property of DP becomes useful for extending word-level perturbations to document-level privatization, in that sequentially performed perturbations can be \textit{composed} to provide document-level guarantees. Concretely, DP composition states that for two DP algorithms $M_1$ with privacy parameter $\varepsilon_1$ and $M_2$ with privacy parameter $\varepsilon_2$, their combination, defined to be $M_{1,2}$: $M_{1,2}$(x) = ($M_1$(x), $M_2$(x)), is ($\varepsilon_1 + \varepsilon_2$)-differentially private. The composition of word-level MLDP to document-level guarantees is illustrated in Figure \ref{fig:w2d}.

The implications of composition in the context of word-level DP are significant. Despite the limitation of single-word perturbations, one can still reason about document-level guarantees by sequentially performing privatization on each word of a given document, and subsequently compose the individual privacy budgets for a composed guarantee \cite{10.1145/3714393.3726504}. This reasoning becomes important in conducting our experiments, where we ensure that every text document is privatized with the same overall document-level budget, an important consideration missed in previous works.

\subsection{Challenges of Word-level DP}
Despite the simplicity and demonstrated effectiveness of word-level MLDP approaches, recent literature has pointed out several shortcomings \cite{mattern-etal-2022-limits}. A clear limitation is the lack of contextual privatization, where word perturbations are performed in isolation, without regard to the surrounding semantic context. Thus, privatized documents achieved via MLDP can lack both grammatical correctness and fluency. In allowing for the release of private \textit{documents}, composition opens up a new danger in the way that privatized documents match the exact length of the original document, due to the one-to-one perturbation, thus already leaking important attributes of the original text and degrading the privacy guarantees offered by DP.

Recent works strive to address some of these challenges, for example by injecting syntax \cite{arnold-etal-2023-guiding} or context \cite{arnold-etal-2023-driving,awon-etal-2025-clusant} into word-level DP. Another recent work by \citet{tong-etal-2025-vulnerability}, which we directly build upon, further investigates privacy vulnerabilities of word-level approaches despite recent advances, where they demonstrate that the contextual remnants after privatization can lead to successful \say{reversals} of word-level perturbations. As previously noted, we indicate several limitations with this work, however, including the sole focus on UMLDP mechanisms and not other MLDP methods, as well as the lack of uniform privacy budgets or use of LLM-based adversarial inference attacks. We design our experiments, outlined in the following, to address these limitations, with the aim of further investigating the potential harms of contextual vulnerability. In the scope of this work, we define \textit{contextual vulnerability} to be semantic remnants of the original, non-privatized texts, which are left due to the randomization inherent to DP processes. These remnants, or \say{clues}, allow for the basis of contextual inference, or inferring the original text (or pieces thereof) based on carried-over context.

\section{Experimental Setup}
We ground our experiments in an investigation of the capabilities of LLMs to infer original semantics given the privatized outputs of word-level MLDP. In following \citet{tong-etal-2025-vulnerability}, we first conduct a case study using the \textsc{SanText} mechanism, exploring the impact of strictly enforcing document-level budgets versus open-ended privatization. Then, we extend our experiments to three recent MLDP mechanisms, with the goal of broadening the generalizability of our findings beyond UMLDP mechanisms. The basis of our experiments, namely the adversarial method, evaluation techniques, and employed metrics, are detailed in the following.

\subsection{Datasets and Privatization}
For DP text sanitization, we use three publicly available datasets. Importantly, these datasets feature user-written texts attributed to distinct authors, and each author in the dataset writes many texts. 

\subsubsection{Datasets}
We choose three datasets that represent sensitive user-written texts, whereby privatization measures would be well-advised. These three datasets are described in the following.

\paragraph{Yelp Reviews}
We use a subset of the Yelp Reviews (YR) corpus as prepared by \citet{utpala-etal-2023-locally}. The dataset contains 17,295 reviews written by the top-10 most frequently writing authors, with the top author associated with 3023 reviews and the tenth author associated with 1391 reviews. The original dataset is intended for binary sentiment analysis, and additionally, we also design an adversarial authorship inference task, mimicking an adversary wishing to infer the identity of an author given only the review text.

\paragraph{Mental Health Blog}
The Mental Health Blog (MHB) dataset \cite{boinepelli-etal-2022-leveraging} contains a large selection of blog posts from an online mental health forum. The posts are categorized by concern, for example \textit{depression} or \textit{anxiety}. From the larger corpus, we select only the posts from the top-50 writing authors, ranging from 115 posts to 8 posts. This subset results in 709 total posts, which are distributed between four categories: \textit{depression}, \textit{anxiety}, \textit{ptsd-trauma}, and \textit{suicidal-thoughts-and-self-harm}. Thus, the dataset presents a four-class classification task, along with a 50-class adversarial inference task.

\paragraph{Enron Emails}
Enron Emails\footnote{\url{https://www.cs.cmu.edu/~enron/}} (EE) is a corpus of about 500k emails from the Enron organization, made public in 2003 during a public investigation carried out by the United States Federal Energy Regulatory Commission. In particular, we use the subset as prepared by \citet{meisenbacher-etal-2025-impact}, which only includes the sent emails from the top-28 users in the corpus, amounting to 12,283 emails, with a maximum of 958 and a minimum of 85 per user. This dataset has no associated downstream task, but we leverage the author IDs for another adversarial authorship inference task.

\subsubsection{Chosen DP Mechanisms}
We first conduct a case study with \textsc{SanText} \cite{yue-etal-2021-differential}, investigating the effect of bounding document-level privacy budgets. Additionally, we experiment with three MLDP mechanisms from the recent literature. These four mechanisms are briefly introduced in the following, where we also refer the interested reader to the original works for further details.

\paragraph{\textsc{SanText} \cite{yue-etal-2021-differential}}
\textsc{SanText} proposes a utility-preserving mechanism for word-level UMLDP, most notably by considering the most \say{suitable} candidates for sanitizing an input word. The private output word, therefore, is produced by randomizing (using the Exponential Mechanism) within this set of candidate replacements. As with \citet{tong-etal-2025-vulnerability}, we experiment with the improved version (\textsc{SanText+}), which goes further to sanitize only the \say{most sensitive} words, which are determined by using frequency statistics from a reference dataset. For the purposes of this work, we use the exact same reference dataset as the original work, namely \textsc{SST-2} \cite{socher-etal-2013-recursive}.

\paragraph{Calibrated Multivariate Perturbations (\textsc{CMP}) \cite{10.1145/3336191.3371856}}
\textsc{CMP}, also kno-wn in the literature as \textsc{MADLIB}, is an early word-level MLDP approach, which calibrates multivariate noise from the normal distribution based on a given input word embedding, adds this noise, and then finds the nearest neighbor to project the perturbed vector back to an output private word. We use a publicly available code implementation \cite{meisenbacher2024comparative}, which leverages 300-dimensional GloVe embeddings \cite{pennington-etal-2014-glove} as the underlying word embedding model. 

\paragraph{Mahalanobis (\textsc{Maha}) \cite{xu2020differentially}}
\textsc{Maha} builds upon \textsc{CMP} by integrating a more advanced distance metric, the Mahalanobis distance, to account for sparse regions in the word embedding space where consistently obtaining perturbed words (beyond the original word) is more difficult. Similar to \textsc{CMP}, we utilize a publicly available code implementation of the \textsc{Maha} mechanism \cite{meisenbacher2024comparative}.

\paragraph{\textsc{1-Diffractor} \cite{10.1145/3643651.3659896}}
\textsc{1-Diffractor} introduces a modified MLDP mechanism, in which word embedding models are projected down to one-dimensional arrays. Calibrated noise is added along this one dimension, allowing for efficient word perturbations. We utilize the implementation made available by the original work, particularly using the \textit{geometric} variant of the mechanism.

\subsubsection{Privatization Approach}
We design a comprehensive experimental approach with respect to sanitization (privatization), arising from the limitations of previous work in the lack of uniform document-level budgets. We tailor document-level privacy budgets ($\varepsilon$ values) to each of our three datasets, and we carefully select base (per-word) $\varepsilon$ values given each of our four chosen mechanisms.

We first begin with a review of the original works proposing our four chosen mechanisms, choosing \textit{base} $\varepsilon$ values, determined by the range of values in the experimental setups of these original works. In particular, we choose three values for each mechanism, representing a strict (low), medium, and lenient (high) $\varepsilon$. These values are shown in Table \ref{tab:epsilons}. Note that for \textsc{CMP} and \textsc{Maha}, we use the same values due to similar values in the original works.

Given the base $\varepsilon$ values, we proceeded to calculate the average number of words in each of the datasets, using the word tokenization tool of the \textsc{nltk} library. Then, document-level budgets were fixed by multiplying each of the base values by the average words result, thereby achieving the per-document privacy budget for all texts in a given dataset (see Table \ref{tab:epsilons}).

With these budgets, we sanitized all texts in each of three datasets, for each of the (\textit{mechanism}, $\varepsilon$) configurations. To accomplish this, the fixed document-level privacy budget was divided by the number of words in a text to be privatized, thereby giving an instance-specific, per-word $\varepsilon$ value. This was done to ensure that all documents (texts) in a dataset achieved an equal (composed) privacy budget, and accordingly, upheld an equal resulting theoretical DP guarantee. Note that for readability, we refer to the \textit{base} $\varepsilon$ when reporting results -- these refer to the document-level budgets for their respective datasets and mechanisms.

\begin{table}[t!]
\scriptsize
\centering
\caption{Chosen \textit{base} and \textit{document} privacy budgets ($\varepsilon$), calculated from the average number of words per document in each dataset. \textsc{CMP} and \textsc{Maha} use the same base $\varepsilon$ values.}
\resizebox{0.7\linewidth}{!}{
\begin{tabular}{lc|ccc}
 &  & YR & MHB & EE \\ \hline
\multicolumn{2}{l|}{Avg. \# words / Base $\varepsilon$} & 208.62 & 304.92 & 77.06 \\ \hline
\multicolumn{1}{l|}{\multirow{3}{*}{\textsc{SanText}}} & 1 &  208 & 304 & 77  \\
\multicolumn{1}{l|}{} & 2 & 417 & 609 & 154  \\
\multicolumn{1}{l|}{} & 3 & 625 & 914 & 231  \\ \hline
\multicolumn{1}{l|}{\multirow{3}{*}{\begin{tabular}[c]{@{}l@{}}\textsc{CMP}\\ \textsc{Maha}\end{tabular}}} & 1 & 208 &  304 &  77 \\
\multicolumn{1}{l|}{} & 10 &  2086 & 3049 & 770 \\
\multicolumn{1}{l|}{} & 20 & 4172 & 6098  &  1541 \\ \hline
\multicolumn{1}{l|}{\multirow{3}{*}{\textsc{1-Diffractor}}} & 0.1 &  20 & 30 & 7 \\
\multicolumn{1}{l|}{} & 1 & 208 & 304 & 77 \\
\multicolumn{1}{l|}{} & 2 & 417 & 609 & 154
\end{tabular}
}
\label{tab:epsilons}
\end{table}

\subsection{LLM-based Contextual Text Reconstruction}
Given all the sanitized counterparts of the original datasets (for all mechanisms and their associated $\varepsilon$ budgets), we proceed to design an LLM-based method for restoring the structure and semantics of the original document given the private document. We ground this approach in the theory that performing word-level MLDP at the document level can leave contextual remnants from the original text, thus degrading the privacy-preserving capabilities of DP text sanitization. This we refer to as \textit{contextual vulnerability}. Therefore, we test the hypothesis that LLMs, with their powerful contextual abilities, pose a risk to word-level MLDP methods, with the potential to \say{undo} the privatization afforded by such methods.

\subsubsection{Attack Vector}
We model an adversary with access to advanced LLMs, as well as with the following information: (1) knowledge of the text sanitization method and the privacy budget used, and (2) knowledge of the domain of the target texts and access to examples resembling these texts.

Given this, the adversary takes the following steps:
\begin{enumerate}
    \itemsep 0em
    \item Curate a few text examples that closely match the target texts. In our experiments, we chose three examples.
    \item Sanitize these texts using the target MLDP method and known privacy budget, yielding (sanitized, original) pairs.
    \item Craft a few-shot LLM prompt with the goal of returning the \say{original} text when a \say{noisy} text is inputted. The prepared (sanitized, original) pairs comprise the few-shot examples.
    \item Given the possession of the sanitized target texts, run these texts through the LLM with the crafted prompt, thereby obtaining reconstructed versions of the sanitized texts.
\end{enumerate}

To create the prompt as described in step (3), we design a prompt to focus on producing a reasonable and coherent \say{clean} output given the \say{noisy} (MLDP-sanitized) input text. We also place an explicit instruction to maintain the same length as the input length, as an adversary would know about the one-to-one mapping of composed word-level perturbations, as well as the fact that the same words should remain the same (i.e., they were not perturbed). The resulting prompt is provided in Table \ref{tab:prompt}.

\begin{table}[t!]
\centering
\scriptsize
\caption{Few-shot Prompt for LLM Text Reconstruction.}
\begin{tabular}{p{0.9\linewidth}}
\hline
\textbf{Prompt}
\\ \hline
You will be given a noisy\_text document.

Your task is to understand the semantic meaning of the text and decrypt the noisy text to its original form.

It is important that the length of the text stays exactly the same. Only replace noisy words with reasonable substitutions. In some cases, words should remain the same.

It is also crucial that the output clean text is coherent and follows a cohesive narrative.
\\ \\
Provide your feedback as follows:

Output:::\\
Clean Text: (your rewritten text)
\\ \\
Here are some examples of how to rewrite noisy texts:
\\ \\
noisy\_text: [SANITIZED TEXT 1]

Output:::
\\Clean Text: [ORIGINAL TEXT 1]
\\ \\
noisy\_text: [SANITIZED TEXT 2]

Output:::
\\Clean Text: [ORIGINAL TEXT 2]
\\ \\
noisy\_text: [SANITIZED TEXT 3]

Output:::
\\Clean Text: [ORIGINAL TEXT 3]
\\ \\
Now here is the noisy text.
\\ \\
noisy\_text: [TARGET TEXT]

Output:::\\
Clean Text:\\ 
\hline
\end{tabular}
\label{tab:prompt}
\end{table}

\subsubsection{Selected LLMs and Prompting Procedure}
To test a variety of currently available LLMs, we select two popular closed-source LLMs and two open-source models. In particular, we choose \textsc{GPT-4o-mini} (2024-07-18) from OpenAI \cite{openai2024gpt4ocard} and \textsc{gemini-2.0-flash} from Google \cite{geminiteam2025geminifamilyhighlycapable}. For open-source models, we use \textsc{Llama-3.3-70B-Instruct-Turbo} \cite{grattafiori2024llama3herdmodels} and \textsc{gemma-3-27b-it} \cite{gemmateam2024gemmaopenmodelsbased}. The closed-source models were called via their respective APIs, and the open-source models were called from the together.ai\footnote{\url{https://www.together.ai/}} platform. Note that we chose to utilize only \say{larger} LLMs due to the assumption that their inferential and generative capabilities would be stronger, and we do not experiment with any smaller LLMs (1B, 3B, etc.). For all LLMs, the default temperature value of 1 was used.

We sequentially prompted (Table \ref{tab:prompt}) the LLMs on all texts in the sanitized dataset variants described previously. This comes with the exception of the first three texts of each dataset, which we held out as the few-shot examples for the prompt. These three texts were not included in any of the subsequent analyses reported in this work, i.e., they are only used for the prompts. After calling the LLMs using the prepared prompts, simple parsing was done to retrieve only the \say{clean text}; no further post-processing was performed. We note that in order to maintain the budget set for our study, the two open-source models were not run on Yelp.

\begin{figure}[t!]
    \centering
    \includegraphics[scale=0.35]{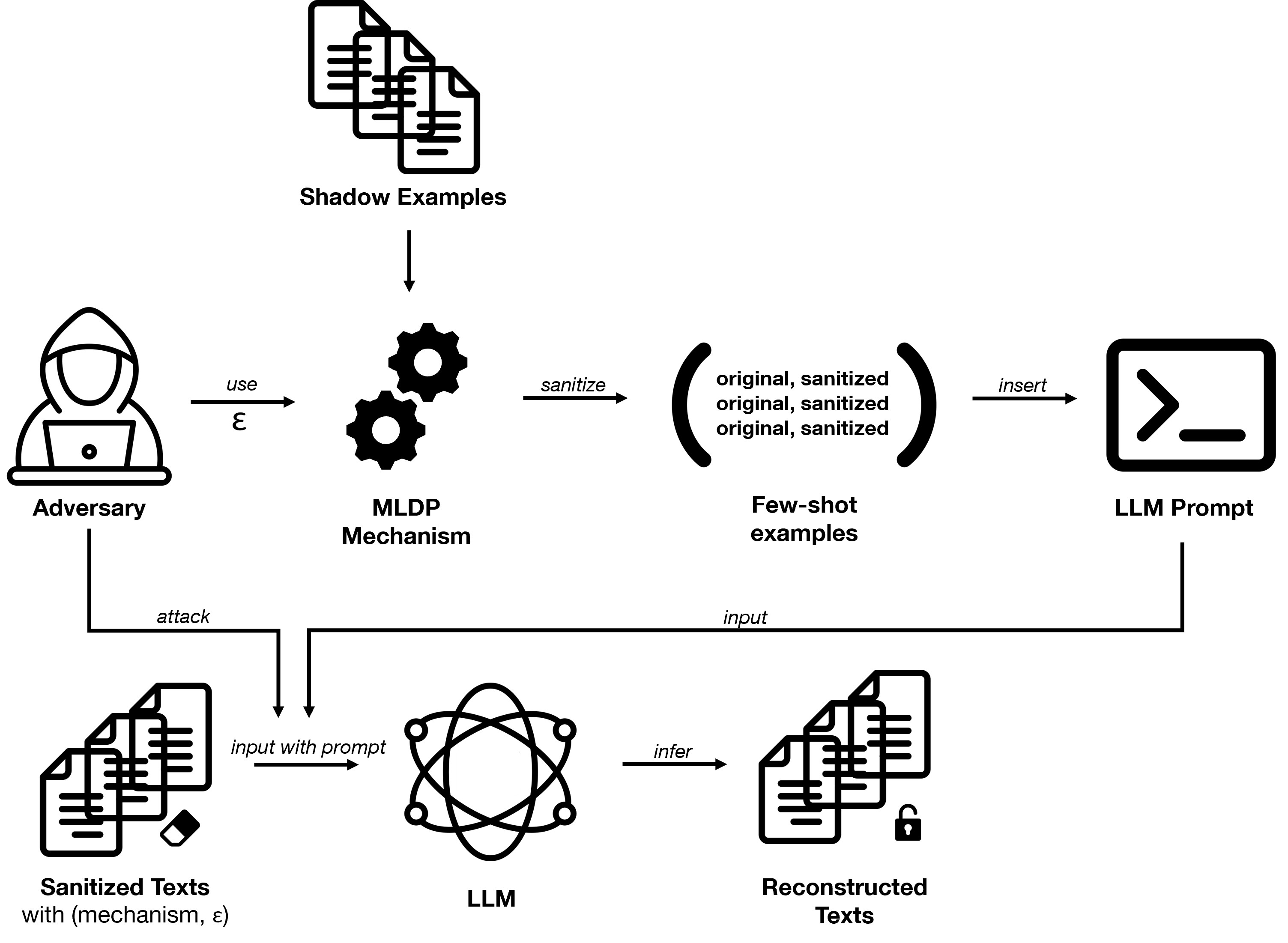}
    \caption{The LLM-based contextual inference process, which aims to reconstruct original texts based on outputs from MLDP mechanisms. The process assumes that an adversary has knowledge of the MLDP mechanism and the utilized privacy budget, and can mimic the sanitization process to produce few-shot examples for the LLM prompt.}
    \label{fig:attack}
\end{figure}

The complete adversarial process is depicted in Figure \ref{fig:attack}.

\subsection{Evaluation Approach}
\label{sec:evaluation}
We conduct a multi-faceted evaluation to investigate the impact on both privacy and utility before and after the LLM reconstruction step. In particular, we explore whether leveraging such an approach can effectively exploit the contextual vulnerability of word-level MLDP, and to what degree. For privacy evaluation, we test the level of defense against adversarial inference, the plausible deniability afforded by the sanitized texts, and the resistance against further contextual inference. We also measure the \textit{utility} of the texts pre- and post-reconstruction, quantified by downstream task performance, semantic similarity, and text coherence. Finally, we calculate the privacy-utility trade-off, observing how this trade-off is affected by sanitization and attempted reversal.

\subsubsection{Privacy Evaluation}
To evaluate privacy, we test two important aspects: defense against inference attacks and plausible deniability (indistinguishability). These are described in the following.

\paragraph{Defense against inference attacks (\textbf{P})}
In this stage of our privacy evaluation, we model an attacker who wishes to obtain the identity of the author of a given text, called \textit{authorship attribution}. Since all of our three datasets contain author information (IDs), such an attack scenario can be modeled. In particular, we assume a 90/10 train/test split, where the train split is in the public domain and the attacker can leverage these texts to train an adversarial classification model to identify an author by their written text (i.e., style, tone, diction, etc.). Then, given the sanitized versions of the test split, the adversary strives to \say{de-anonymize} the respective authors using the trained classification model.

Following the recent literature \cite{mattern-etal-2022-limits,utpala-etal-2023-locally}, we model two variants of the above-described attacker. The \textit{static} (s) attacker does not have knowledge of the sanitization method or privacy budget, and therefore must train the attack model on the original (clean) texts (train split). The more capable \textit{adaptive} (a) attacker does indeed have this knowledge, and uses this knowledge to sanitize the train split, in order to better mirror the target test split. Following the training of their adversarial classification models, both attackers evaluate their models on the sanitized test split. For the classification models, we fine-tune \textsc{deberta-v3-base} \cite{he2021deberta} models on the respective train sets for one epoch (YR and EE) or three epochs (MHB), using all default parameters of the HuggingFace Trainer toolkit. The choice of three epochs was made due to the smaller size of MHB.

To measure the defense afforded by sanitization methods against these attackers, we first measure the performance of the static attacker on the \textit{non-sanitized} test splits of the three datasets, which represent the baseline. Then, all subsequent evaluations on the sanitized test splits are compared against the baseline, and the difference in adversarial performance represents the empirical privacy gains afforded by privatization. For all tests, we use the micro-F1 score from the authorship attribution task as the performance metric.

\paragraph{Plausible deniability (indistinguishability, \textbf{In})}
At the basis of all DP mechanisms is the injection of \textit{plausible deniability} into a dataset, or rather, into computations performed on the data. At their core, DP mechanisms (including those in NLP) strive to offer plausible deniability by making it difficult to deduce what the input to a computation or query was, given only the noisy output result. In the context of MLDP, a good mechanism should yield such plausible deniability in the form that a private output (document) is reasonably \textit{indistinguishable} from its original input.

We measure plausible deniability by introducing an \textit{indistinguishability} metric, which provides an idea of how closely and to what degree, on average, the private output text resembles its original input. Intuitively, a mechanism that always outputs private texts that are nearly similar to the inputs does not grant plausible deniability, as any adversary could assume that much of the original semantics has been preserved. Conversely, when a private output text is often very distant from the original, an adversary could not say with certainty what the original text content could have been.

The \textit{indistinguishability} metric is measured by first calculating the \textit{sentence embeddings} of each document in the private output dataset. We perform this using Sentence Transformers \cite{reimers-gurevych-2019-sentence}, specifically the \textsc{all-MiniLM-L12-v2} model\footnote{\url{https://huggingface.co/sentence-transformers/all-MiniLM-L12-v2}}. Then, in an iterative fashion, each text from the original (non-sanitized) dataset counterpart is embedded, and a nearest neighbor search is performed to calculate at which $k$ the true private counterpart to the original text is the $k$th nearest neighbor. The $k$ for each (original, private) pair is averaged for the whole dataset, and the resulting \textit{indistinguishability} score is represented as a percentage of the size of the dataset. For example, in a dataset with 100 texts, a score of 0 would imply that the private text is always the closest semantically to the original (the nearest neighbor), whereas a score of 1 would imply strong indistinguishability, as the original and private texts are always semantically distant (100th nearest neighbor).

To accommodate the different dataset sizes, we set the max $k$ value to be the size of each respective dataset. For nearest neighbor search, we utilize \textsc{semantic\_search} from Sentence Transformers.



% ...

\subsubsection{Utility Evaluation}
The utility evaluation consists of downstream task performance measurement (in the case of YR and MHB), as well as the calculation of semantic similarity and coherence of the sanitized texts. These evaluations paint a picture of the impact of sanitization on the usefulness of the text, and they also enable the calculation of the privacy-utility trade-off.

\paragraph{Downstream task performance (\textbf{Util})}
Since both YR and MHB contain an associated downstream task (sentiment analysis and ailment detection, respectively), we are able to measure the effect of text sanitization on downstream task performance. We fine-tune a \textsc{deberta-v3-base} model on a 90\% train split of all variants of the YR and MHB datasets, for one and three epochs, respectively. In particular, YR is used for a binary sentiment classification and MHB for a four-class classification task. The micro-F1 score on the 10\% test split is recorded (due to class imbalance), both for the baseline (non-sanitized) and sanitized datasets. Each training and evaluation procedure is repeated three times to account for variances in training, and the reported scores represent the average.

\paragraph{Semantic similarity (\textbf{SS})}
This metric refers to a representation of the similarity in meaning of two given texts. Typically, semantic similarity is measured by calculating the embeddings (vector representations) of two texts, and then measuring the \textit{cosine similarity} between the two embeddings. Here, a score of 1 means perfect similarity, where a score of -1 denotes very distant meanings.

As with \textit{indistinguishability}, we use Sentence Transformers with an \textsc{all-MiniLM-L12-v2} model to compute the embeddings for all sanitized texts and their original counterparts. Then, we calculate the cosine similarity between all of these pairs, averaging the results for a single semantic similarity score per private dataset. Thus, we measure for each sanitization setting (dataset, mechanism, budget) the average semantic similarity over all texts within the corresponding dataset. This provides a broad overview of the preservation of semantics (meaning), which lends to the utility of a dataset.

\paragraph{Text coherence (\textbf{Co})}
A known limitation of word-level MLDP methods is the tendency to produce output texts that are lacking in coherence or fluency, and moreover, that contain grammatical incorrectness or inconsistencies. As a proxy to capture the preservation of such coherence, we use the \textit{perplexity} metric. In essence, perplexity mirrors text coherence by measuring how \say{surprised} a language model is to see the next token over the tokens of a text. High perplexity indicates incorrect, inconsistent, or non-fluent writing, whereas low perplexity denotes that the text aligns well with the language model's understanding of proper text.

We measure the mean perplexity of all texts in our private datasets, i.e., the average of the perplexity scores for each sanitized text in the dataset. As perplexity is \textit{reference-free} (does not require comparison to a reference text), the original text counterpart is not taken into consideration. As the underlying language model, we use the \textsc{GPT-2} model \cite{radford2019language}, following related works in DP NLP that also use perplexity in their evaluation \cite{10.1145/3485447.3512232, 10.1145/3664476.3669926}.

\subsubsection{Quantifying the privacy-utility trade-off}
An important consideration in the evaluation of any text privatization method, especially those leveraging DP, comes with the relative gains of doing so, with respect to the potential utility losses incurred. This is often referred to as the \textit{privacy-utility trade-off}, and it can be calculated by directly weighing privacy gains against utility losses.

Due to the direct comparability, we measure the observable trade-offs when considering the \textit{defense against inference attacks} and \textit{downstream task performance}. We define $\mathcal{U}_o$ to be the baseline utility score, which is represented as the downstream task performance using the original, non-sanitized dataset. We also define $\mathcal{U}_p$ to be the same utility score (F1) measured on the private datasets. Similarly, we define $P_o$ and $P_p$ as the privacy scores, namely the adversarial performance on the baseline and private datasets, respectively. We thus define the privacy-utility trade-off as: $\mathbf{TO} = \frac{\mathcal{U}_p}{\mathcal{U}_o} - \frac{P_p}{P_o}$. In the case of the $\mathcal{U}$ calculation for Yelp, we account for the highly imbalanced dataset (towards positive reviews) by considering the utility change \textit{over majority-class guessing} (MG), thus yielding $\mathcal{U} = \mathcal{U}_{observed} - \mathcal{U}_{MG}$, for both $\mathcal{U}_o$ and $\mathcal{U}_p$.

A positive $TO$ score indicates that the privacy benefits of sanitization outweigh utility losses, whereas a negative score would imply the opposite. The strength (magnitude) of this score suggests how clear the trade-off is. Since we use the \textit{defense against inference} scores, we calculate $TO$ for both the static and adaptive settings.

\section{Experiment Results}
We present the results of our experiments, beginning with a case study of the \textsc{SanText} mechanism, which reveals the impact of imposing document-level budgets during sanitization. Then, we present the privacy and utility results of the remaining three mechanisms, both before and after LLM-based reconstruction.

\subsection{Enforcing Uniform Document-level MLDP: A Case Study with \textsc{SanText}}
\label{sec:case}
As introduced, we initiate our investigation with a case study of the \textsc{SanText} mechanism, particularly to determine the impact that bounding document-level privacy budgets has, in comparison to previous work which uses unbounded sanitization \cite{tong-etal-2025-vulnerability}. The findings of this investigation serve to guide the remainder of this work, where we evaluate the remaining three MLDP mechanisms.

\begin{table}[t!]
\scriptsize
\centering
\caption{Results of the \textsc{SanText} case study, where we directly compare the effects of enforcing uniform (bounded) document-level privacy budgets or not. Where applicable (\textit{Util} and \textit{P)}, we present the average F1 score (over three runs), with the std. dev. as a subscript. We refer the reader to Section \ref{sec:evaluation} for descriptions of the evaluation metrics. $\varepsilon$ refers to the \textit{base} (per-word) privacy budgets. $\uparrow$ = higher scores are better.}
    \begin{subtable}[t]{0.95\linewidth}
    \scriptsize
     \centering
        \caption{YR}
\resizebox{0.9\linewidth}{!}{
\begin{tabular}{ll|ccc|ccc}
&  & \multicolumn{3}{c|}{Unbounded} & \multicolumn{3}{c}{Bounded} \\ \cline{3-8} 
 & \multicolumn{1}{r|}{$\varepsilon$} & 1 & 2 & 3 & 1 & 2 & 3 \\ \hline
\multicolumn{1}{l|}{\multirow{2}{*}{Original}} & Util & \multicolumn{6}{c}{$95.68_{0.2}$} \\
\multicolumn{1}{l|}{} & P & \multicolumn{6}{c}{95.03} \\ \hline
\multicolumn{1}{l|}{\multirow{8}{*}{\textsc{SanText}}} & Util $\uparrow$ & $93.2_{0.0}$ & $93.2_{0.0}$ & $93.8_{0.8}$ & $93.2_{0.0}$ & $93.2_{0.0}$ & $93.2_{0.0}$ \\
\multicolumn{1}{l|}{} & SS $\uparrow$  & 0.398 & 0.448 & 0.501 & 0.421 & 0.470 & 0.503 \\
\multicolumn{1}{l|}{} & Co $\downarrow$ & 1004 & 1041 & 997 & 1009 & 1001 & 970 \\
\multicolumn{1}{l|}{} & P(s) $\downarrow$ & 41.68 & 43.93 & 45.09 & 43.24 & 43.70 & 45.72 \\
\multicolumn{1}{l|}{} & P(a) $\downarrow$ & $84.8_{1.6}$ & $85.9_{0.3}$ & $85.6_{1.8}$ & $82.9_{2.2}$ & $86.2_{1.6}$ & $82.6_{1.7}$ \\
\multicolumn{1}{l|}{} & In $\uparrow$  & 0.088 & 0.059 & 0.041 & 0.079 & 0.063 & 0.052 \\
\multicolumn{1}{l|}{} & TO(s) $\uparrow$ & 0.53 & 0.51 & 0.51 & 0.52 & 0.51 & 0.49  \\
\multicolumn{1}{l|}{} & TO(a) $\uparrow$ & 0.08 & 0.07 & 0.08 & 0.10 & 0.07 & 0.10 \\ \hline
\multicolumn{1}{l|}{\multirow{8}{*}{\textsc{GPT}}} & Util $\uparrow$ & $93.8_{0.8}$ & $94.6_{1.1}$ & $93.9_{0.8}$ & $95.3_{0.2}$ & $94.7_{0.4}$ & $94.7_{1.0}$ \\
\multicolumn{1}{l|}{} & SS $\uparrow$  & 0.544 & 0.636 & 0.665 & 0.615 & 0.648 & 0.673 \\
\multicolumn{1}{l|}{} & Co $\downarrow$ & 83 & 56 & 47 & 50 & 50 & 45 \\
\multicolumn{1}{l|}{} & P(s) $\downarrow$ & 32.14 & 36.94 & 34.74 & 32.83 & 35.09 & 35.49 \\
\multicolumn{1}{l|}{} & P(a) $\downarrow$ & $73.9_{1.1}$ & $72.8_{0.4}$ & $76.6_{0.2}$ & $68.6_{3.0}$ & $71.2_{2.6}$ & $72.3_{0.4}$ \\
\multicolumn{1}{l|}{} & In $\uparrow$  & 0.077 & 0.026 & 0.022 & 0.033 & 0.023 & 0.015 \\
\multicolumn{1}{l|}{} & TO(s) $\uparrow$ & \textbf{0.64} & \textbf{0.58} & \textbf{0.61} & \textbf{0.63} & \textbf{0.60} & \textbf{0.60} \\
\multicolumn{1}{l|}{} & TO(a) $\uparrow$ & \textbf{0.20} & \textbf{0.21} & \textbf{0.17 }& \textbf{0.25} & \textbf{0.22} & \textbf{0.21} \\ \hline
\multicolumn{1}{l|}{\multirow{8}{*}{\textsc{gemini}}} & Util $\uparrow$ & $93.2_{0.6}$ & $93.6_{0.7}$ & $92.7_{0.0}$ & $92.8_{0.2}$ & $93.1_{0.6}$ & $94.2_{0.1}$ \\
\multicolumn{1}{l|}{} & SS $\uparrow$  & 0.499 & 0.611 & 0.644 & 0.586 & 0.580 & 0.625 \\
\multicolumn{1}{l|}{} & Co $\downarrow$ & 669 & 456 & 388 & 471 & 623 & 508 \\
\multicolumn{1}{l|}{} & P(s) $\downarrow$ & 46.36 & 59.42 & 59.42 & 57.63 & 54.34 & 56.42 \\
\multicolumn{1}{l|}{} & P(a) $\downarrow$ & $80.4_{1.1}$ & $82.3_{1.8}$ & $83.6_{0.4}$ & $79.2_{5.1}$ & $82.0_{2.3}$ & $84.0_{0.3}$ \\
\multicolumn{1}{l|}{} & In $\uparrow$  & 0.089 & 0.046 & 0.030 & 0.056 & 0.051 & 0.037 \\
\multicolumn{1}{l|}{} & TO(s) $\uparrow$ & 0.49 & 0.35 & 0.35 & 0.37 & 0.40 & 0.38  \\
\multicolumn{1}{l|}{} & TO(a) $\uparrow$ & 0.13 & 0.11 & 0.10 & 0.14 & 0.11 & 0.09 \\
\end{tabular}
}
    \end{subtable}     
    \vspace{3pt}
    
    \begin{subtable}[t]{0.95\linewidth}
    \scriptsize
     \centering
        \caption{MHB}
\resizebox{0.9\linewidth}{!}{
\begin{tabular}{ll|ccc|ccc}
&  & \multicolumn{3}{c|}{Unbounded} & \multicolumn{3}{c}{Bounded} \\ \cline{3-8} 
 & \multicolumn{1}{r|}{$\varepsilon$} & 1 & 2 & 3 & 1 & 2 & 3 \\ \hline
\multicolumn{1}{l|}{\multirow{2}{*}{Original}} & Util & \multicolumn{6}{c}{$71.83_{1.1}$} \\
\multicolumn{1}{l|}{} & P & \multicolumn{6}{c}{23.94} \\ \hline
\multicolumn{1}{l|}{\multirow{8}{*}{\textsc{SanText}}} & Util $\uparrow$ & $63.4_{0.0}$ & $61.0_{3.3}$ & $63.4_{0.0}$ & $64.8_{2.0}$ & $63.4_{0.0}$ & $63.8_{1.8}$ \\
\multicolumn{1}{l|}{} & SS $\uparrow$  & 0.410 & 0.456 & 0.515 & 0.453 & 0.507 & 0.531 \\
\multicolumn{1}{l|}{} & Co $\downarrow$ & 850 & 859 & 830 & 805 & 952 & 769 \\
\multicolumn{1}{l|}{} & P(s) $\downarrow$ & 19.72 & 23.94 & 19.72 & 15.49 & 18.31 & 19.72 \\
\multicolumn{1}{l|}{} & P(a) $\downarrow$ & $23.9_{0.0}$ & $23.9_{0.0}$ & $23.9_{0.0}$ & $23.9_{0.0}$ & $23.9_{0.0}$ & $23.9_{0.0}$ \\
\multicolumn{1}{l|}{} & In $\uparrow$  & 0.093 & 0.070 & 0.052 & 0.075 & 0.061 & 0.049 \\
\multicolumn{1}{l|}{} & TO(s) $\uparrow$ & \textbf{0.06} & \textbf{-0.15} & \textbf{0.06} & \textbf{0.25} & \textbf{0.12} & \textbf{0.06}  \\
\multicolumn{1}{l|}{} & TO(a) $\uparrow$ & \textbf{-0.12} & \textbf{-0.15} & \textbf{-0.12 }& \textbf{-0.10} & \textbf{-0.12} & \textbf{-0.11} \\ \hline
\multicolumn{1}{l|}{\multirow{8}{*}{\textsc{GPT}}} & Util $\uparrow$ & $54.9_{1.1}$ & $54.0_{0.7}$ & $53.1_{2.9}$ & $51.6_{1.3}$ & $54.5_{1.3}$ & $56.3_{4.0}$ \\
\multicolumn{1}{l|}{} & SS $\uparrow$  & 0.584 & 0.616 & 0.717 & 0.684 & 0.705 & 0.718 \\
\multicolumn{1}{l|}{} & Co $\downarrow$ & 58 & 69 & 35 & 48 & 36 & 35 \\
\multicolumn{1}{l|}{} & P(s) $\downarrow$ & 25.35 & 23.94 & 23.94 & 23.94 & 23.94 & 23.94 \\
\multicolumn{1}{l|}{} & P(a) $\downarrow$ & $23.9_{0.0}$ & $23.9_{0.0}$ & $23.9_{0.0}$ & $23.9_{0.0}$ & $23.9_{0.0}$ & $23.9_{0.0}$ \\
\multicolumn{1}{l|}{} & In $\uparrow$  & 0.059 & 0.046 & 0.017 & 0.018 & 0.010 & 0.011 \\
\multicolumn{1}{l|}{} & TO(s) $\uparrow$ & -0.18 & \textbf{-0.15} & -0.12 & -0.10 & -0.12 & -0.11  \\
\multicolumn{1}{l|}{} & TO(a) $\uparrow$ & \textbf{-0.12} & \textbf{-0.15} & \textbf{-0.12} & \textbf{-0.10} & \textbf{-0.12} & \textbf{-0.11} \\ \hline
\multicolumn{1}{l|}{\multirow{8}{*}{\textsc{gemini}}} & Util $\uparrow$ & $51.6_{3.7}$ & $52.1_{2.0}$ & $54.0_{0.7}$ & $49.3_{4.1}$ & $54.5_{1.3}$ & $53.5_{0.0}$ \\
\multicolumn{1}{l|}{} & SS $\uparrow$  & 0.542 & 0.625 & 0.694 & 0.683 & 0.699 & 0.686 \\
\multicolumn{1}{l|}{} & Co $\downarrow$ & 532 & 368 & 315 & 270 & 459 & 363 \\
\multicolumn{1}{l|}{} & P(s) $\downarrow$ & 22.54 & 26.76 & 22.54 & 22.54 & 19.72 & 21.13 \\
\multicolumn{1}{l|}{} & P(a) $\downarrow$ & $23.9_{0.0}$ & $23.9_{0.0}$ & $23.9_{0.0}$ & $23.9_{0.0}$ & $23.9_{0.0}$ & $23.9_{0.0}$ \\
\multicolumn{1}{l|}{} & In $\uparrow$  & 0.088 & 0.056 & 0.043 & 0.032 & 0.025 & 0.032 \\
\multicolumn{1}{l|}{} & TO(s) $\uparrow$ & -0.18 & \textbf{-0.15} & -0.12 & -0.10 & -0.12 & -0.11 \\
\multicolumn{1}{l|}{} & TO(a) $\uparrow$ & \textbf{-0.12} & \textbf{-0.15} & \textbf{-0.12} & \textbf{-0.10} & \textbf{-0.12} & \textbf{-0.11} \\
\end{tabular}
}
    \end{subtable}

    \vspace{3pt}
    
    \begin{subtable}[t]{0.95\linewidth}
    \scriptsize
     \centering
        \caption{EE}
\resizebox{0.9\linewidth}{!}{
\begin{tabular}{ll|ccc|ccc}
&  & \multicolumn{3}{c|}{Unbounded} & \multicolumn{3}{c}{Bounded} \\ \cline{3-8} 
 & \multicolumn{1}{r|}{$\varepsilon$} & 1 & 2 & 3 & 1 & 2 & 3 \\ \hline
\multicolumn{1}{l|}{Original} & P & \multicolumn{6}{c}{38.03} \\ \hline
\multicolumn{1}{l|}{\multirow{5}{*}{\textsc{SanText}}} & SS $\uparrow$  & 0.341 & 0.405 & 0.474 & 0.422 & 0.477 & 0.504 \\
\multicolumn{1}{l|}{} & Co $\downarrow$ & 1903 & 1902 & 1970 & 1853 & 1755 & 1708\\
\multicolumn{1}{l|}{} & P(s) $\downarrow$ & 5.70 & 6.60 & 6.11 & 7.33 & 6.92 & 7.74 \\
\multicolumn{1}{l|}{} & P(a) $\downarrow$ & $15.3_{0.9}$ & $16.3_{1.6}$ & $16.5_{3.7}$ & $16.5_{3.2}$ & $17.5_{4.0}$ & $19.6_{1.7}$ \\
\multicolumn{1}{l|}{} & In $\uparrow$  & 0.075 & 0.050 & 0.029 & 0.052 & 0.035 & 0.028 \\ \hline
\multicolumn{1}{l|}{\multirow{5}{*}{\textsc{GPT}}} & SS $\uparrow$  & 0.439 & 0.480 & 0.546 & 0.499 & 0.550 & 0.573 \\
\multicolumn{1}{l|}{} & Co $\downarrow$ & 198 & 190 & 233 & 139 & 131 & 142\\
\multicolumn{1}{l|}{} & P(s) $\downarrow$ & 18.89 & 20.85 & 19.46 & 17.59 & 21.01 & 21.74 \\
\multicolumn{1}{l|}{} & P(a) $\downarrow$ & $17.8_{1.7}$ & $19.8_{0.2}$ & $19.8_{0.3}$ & $17.5_{0.7}$ & $22.2_{0.8}$ & $21.9_{0.9}$ \\
\multicolumn{1}{l|}{} & In $\uparrow$  & 0.074 & 0.056 & 0.038 & 0.050 & 0.032 & 0.024 \\ \hline
\multicolumn{1}{l|}{\multirow{5}{*}{\textsc{gemini}}} & SS $\uparrow$  & 0.378 & 0.428 & 0.515 & 0.431 & 0.502 & 0.518 \\
\multicolumn{1}{l|}{} & Co $\downarrow$ & 1665 & 1666 & 1433 & 1787 & 1571 & 1619 \\
\multicolumn{1}{l|}{} & P(s) $\downarrow$ & 9.20 & 9.12 & 11.64 & 7.98 & 9.20 & 9.61 \\
\multicolumn{1}{l|}{} & P(a) $\downarrow$ & $11.9_{3.6}$ & $13.5_{1.1}$ & $16.4_{0.5}$ & $13.2_{2.0}$ & $18.4_{1.8}$ & $17.8_{1.2}$ \\
\multicolumn{1}{l|}{} & In $\uparrow$ & 0.079 & 0.052 & 0.031 & 0.052 & 0.034 & 0.027 \\
\end{tabular}
}
    \end{subtable}
\label{tab:results_santext}
\end{table}

\subsubsection{Insights from bounding document budgets}
The full results of the \textsc{SanText} case study are presented in Table \ref{tab:results_santext}. In first analyzing the raw \textsc{SanText} results (pre-LLM), we find that in some cases, fixing document budgets (rather than unbounded per-word budgets) leads to higher utility, but lower privacy. Relatedly, the bounded results can also produce private outputs that are more semantically reminiscent of the original texts, while also producing more coherent texts. As a result, the bounded results consistently lead to better trade-offs (as exhibited in the case of YR and MHB), showing a potential benefit of ensuring uniform privacy budgets. We caution, however, that this is not always the case, and in some direct comparisons, the results are mixed (e.g., SS and Co for YR).

The data reconstruction process with the selected LLMs seems to amplify some of these disparities, yet those resulting from bounded privatization consistently show superior results in terms of semantic similarity and coherence. However, we observe in many cases that defense against inference attacks (both static and adaptive) is weaker in the bounded privatization results, suggesting that this may lead to a wider attack vector than the unbounded case.

We find a potential reason for the mixed results: the \textit{variability} of document lengths. Since all of the texts in a dataset are of various lengths, fixing document-level privacy budgets to the \textit{average} number of words will result in very long documents being privatized particularly strictly, and rather short documents very leniently. While this is an inherent limitation of LDP across a dataset, the results of Table \ref{tab:results_santext} suggest that certain benefits do realize, perhaps outweighing the downsides of \say{incongruous} privatization. Above all, however, the bounding of document-level privacy budgets is \textit{necessary} in the uniform and fair evaluation of MLDP mechanisms across texts and datasets, where unbounded privatization leads to different guarantees across texts of different length (in the case of composed MLDP guarantees).

\subsubsection{Takeaways for our work}
Learning from the \textsc{SanText} case study, we continue with the investigation of the three other selected MLDP mechanisms, namely by using them in a \textit{bounded} fashion. Beyond ensuring a theoretically correct and fair evaluation, we wish to observe whether the same trends surface when studying the transition from original to private to reconstructed text. This decision to enforce document-level budgets is particularly supported by the observation of the resulting trade-off scores, where the bounded results nearly always exhibit similar or better trade-offs than unbounded privatization.

From the case study, we also learn of the importance of \textit{dataset} and \textit{LLM} in drawing conclusions about the effectiveness of the LLM contextual inference. As an example, LLMs perform much better (namely, in recovering semantics) in the YR tasks than in MHB, and in YR, \textsc{GPT-4o} displays a clear advantage over \textsc{Gemini-2.0}. These results demonstrate the importance of fundamental design choices when setting up adversarial tasks, and we validate these discrepancies in studying the three further MLDP methods.

\subsection{How vulnerable are MLDP methods, and to what effect?}
Focusing on our three selected MLDP text sanitization methods, we systematically report and analyze the results of our conducted experiments, which follow the same structure as those in the \textsc{SanText} case study. These results are reported in full in Table \ref{tab:results_mldp}.

\begin{table}[htbp]
\scriptsize
\centering
\caption{Results of the privacy and utility experiments involving the three selected MLDP mechanisms, the three selected datasets, and all privacy budget settings. For each (\textit{mechanism}, $\varepsilon$) pair, we bold the highest scoring trade-off (TO) value, for both the static and adaptive attacker settings. 
}
    \begin{subtable}[t]{0.98\linewidth}
    \scriptsize
     \centering
        \caption{YR}
\resizebox{0.97\linewidth}{!}{
\begin{tabular}{ll|ccc|ccc|ccc}
&  & \multicolumn{3}{c|}{\textsc{CMP}} & \multicolumn{3}{c|}{\textsc{Maha}} & \multicolumn{3}{c}{\textsc{1-Diffractor}} \\ \cline{3-11} 
 & \multicolumn{1}{r|}{$\varepsilon$} & 1 & 10 & 20 & 1 & 10 & 20 & 0.1 & 1 & 2 \\ \hline
\multicolumn{1}{l|}{\multirow{2}{*}{Original}} & Util & \multicolumn{9}{c}{$95.68_{0.2}$} \\
\multicolumn{1}{l|}{} & P & \multicolumn{9}{c}{95.03} \\ \hline
\multicolumn{1}{l|}{\multirow{8}{*}{\textsc{MLDP}}} & Util $\uparrow$ & $93.2_{0.0}$ & $93.3_{0.2}$ & $93.2_{0.0}$ & $93.2_{0.0}$ & $93.2_{0.0}$ & $93.2_{0.0}$ & $93.2_{0.0}$ & $94.1_{0.9}$ & $93.5_{0.4}$ \\
\multicolumn{1}{l|}{} & SS $\uparrow$  & 0.168 & 0.453 & 0.710 & 0.167 & 0.397 & 0.639 & 0.511 & 0.855 & 0.930 \\
\multicolumn{1}{l|}{} & Co $\downarrow$ & 1964 & 3130 & 1636 & 2104 & 3071 & 1862 & 619 & 179 & 108 \\
\multicolumn{1}{l|}{} & P(s) $\downarrow$ & 20.23 & 34.10 & 55.90 & 20.93 & 31.16 & 50.29 & 46.88 & 70.98 & 81.56 \\
\multicolumn{1}{l|}{} & P(a) $\downarrow$ & $73.5_{0.3}$ & $76.0_{1.4}$ & $84.0_{1.0}$ & $73.3_{1.6}$ & $74.0_{1.3}$ & $80.5_{0.9}$ & $88.4_{0.8}$ & $92.3_{0.2}$ & $93.9_{0.6}$ \\
\multicolumn{1}{l|}{} & In $\uparrow$ & 0.343 & 0.187 & 0.081 & 0.346 & 0.206 & 0.104 & 0.058 & 0.001 & 0.000 \\
\multicolumn{1}{l|}{} & TO(s) $\uparrow$ & 0.76 & 0.62 & 0.39 & 0.75 & 0.65 & 0.44 & \textbf{0.48} & \textbf{0.24} & 0.12 \\
\multicolumn{1}{l|}{} & TO(a) $\uparrow$ & 0.20 & 0.18 & 0.09 & 0.20 & 0.19 & 0.13 & 0.04 & 0.01 & -0.01 \\ \hline
\multicolumn{1}{l|}{\multirow{8}{*}{\textsc{GPT}}} & Util $\uparrow$ & $92.7_{0.0}$ & $92.7_{0.0}$ & $92.7_{0.0}$ & $92.7_{0.0}$ & $93.0_{0.4}$ & $94.2_{1.0}$ & $93.5_{0.8}$ & $94.6_{1.4}$ & $93.0_{0.5}$ \\
\multicolumn{1}{l|}{} & SS $\uparrow$  & 0.236 & 0.526 & 0.757 & 0.240 & 0.481 & 0.703 & 0.648 & 0.883 & 0.938  \\
\multicolumn{1}{l|}{} & Co $\downarrow$ & 276 & 757 & 572 & 254 & 699 & 584 & 60 & 47 & 44 \\
\multicolumn{1}{l|}{} & P(s) $\downarrow$ & 11.27 & 32.08 & 52.49 & 11.21 & 28.32 & 47.28 & 59.13 & 71.97 & 74.57 \\
\multicolumn{1}{l|}{} & P(a) $\downarrow$ & $25.6_{0.9}$ & $50.6_{0.9}$ & $76.5_{0.4}$ & $26.4_{0.9}$ & $51.7_{1.7}$ & $70.5_{1.2}$ & $80.8_{3.0}$ & $88.3_{2.3}$ & $89.2_{2.9}$ \\
\multicolumn{1}{l|}{} & In $\uparrow$ & 0.396 & 0.185 & 0.068 & 0.393 & 0.210 & 0.089 & 0.020 & 0.000 & 0.000 \\
\multicolumn{1}{l|}{} & TO(s) $\uparrow$ & \textbf{0.85} & \textbf{0.64} & \textbf{0.42} & \textbf{0.86} & \textbf{0.68} & \textbf{0.48} & 0.35 & 0.23 & \textbf{0.19} \\
\multicolumn{1}{l|}{} & TO(a) $\uparrow$ & \textbf{0.70 }& \textbf{0.44 }& \textbf{0.17} & \textbf{0.70} & \textbf{0.43 }& 0.23 & \textbf{0.12} & \textbf{0.05 }& \textbf{0.04 }\\ \hline
\multicolumn{1}{l|}{\multirow{8}{*}{\textsc{gemini}}} & Util $\uparrow$ & $93.5_{0.0}$ & $92.7_{0.0}$ & $92.7_{0.0}$ & $92.7_{0.0}$ & $92.8_{0.0}$ & $92.7_{0.0}$ & $93.0_{0.4}$ & $94.4_{1.2}$ & $94.6_{1.5}$ \\
\multicolumn{1}{l|}{} & SS $\uparrow$  & 0.257 & 0.490 & 0.746 & 0.275 & 0.405 & 0.651 & 0.618 & 0.885 & 0.942 \\
\multicolumn{1}{l|}{} & Co $\downarrow$ & 497 & 1751 & 993 & 574 & 2680 & 1653 & 357 & 103 & 80 \\
\multicolumn{1}{l|}{} & P(s) $\downarrow$ & 23.53 & 34.34 & 59.02 & 21.33 & 31.16 & 50.64 & 61.79 & 84.05 & 86.36 \\
\multicolumn{1}{l|}{} & P(a) $\downarrow$ & $34.7_{1.0}$ & $56.9_{1.9}$ & $81.7_{1.0}$ & $34.3_{0.5}$ & $65.1_{1.3}$ & $69.9_{2.1}$ & $89.5_{0.7}$ & $91.6_{1.5}$ & $93.9_{0.3}$ \\
\multicolumn{1}{l|}{} & In $\uparrow$ & 0.411 & 0.201 & 0.076 & 0.396 & 0.211 & 0.101 & 0.029 & 0.000 & 0.000 \\
\multicolumn{1}{l|}{} & TO(s) $\uparrow$ & 0.73 & 0.61 & 0.35 & 0.75 & 0.65 & 0.44 & 0.32 & 0.10 & 0.07 \\
\multicolumn{1}{l|}{} & TO(a) $\uparrow$ & 0.61 & 0.38 & 0.11 & 0.61 & 0.29 & \textbf{0.24} & 0.03 & 0.02 & -0.01 \\ \hline
\end{tabular}
}
    \end{subtable}
    \vspace{5pt}
    
    \begin{subtable}[t]{0.98\linewidth}
    \scriptsize
     \centering
        \caption{MHB}
\resizebox{0.97\linewidth}{!}{
\begin{tabular}{ll|ccc|ccc|ccc}
&  & \multicolumn{3}{c|}{\textsc{CMP}} & \multicolumn{3}{c|}{\textsc{Maha}} & \multicolumn{3}{c}{\textsc{1-Diffractor}} \\ \cline{3-11} 
 & \multicolumn{1}{r|}{$\varepsilon$} & 1 & 10 & 20 & 1 & 10 & 20 & 0.1 & 1 & 2 \\ \hline
\multicolumn{1}{l|}{\multirow{2}{*}{Original}} & Util & \multicolumn{9}{c}{$71.83_{1.1}$} \\
\multicolumn{1}{l|}{} & P & \multicolumn{9}{c}{23.94} \\ \hline
\multicolumn{1}{l|}{\multirow{8}{*}{\textsc{MLDP}}} & Util $\uparrow$ & $63.4_{0.0}$ & $63.4_{1.1}$ & $68.5_{0.7}$ & $63.4_{0.0}$ & $65.7_{2.4}$ & $71.4_{0.7}$ & $65.3_{3.7}$ & $69.5_{2.4}$ & $73.2_{1.1}$ \\
\multicolumn{1}{l|}{} & SS $\uparrow$  & 0.079 & 0.427 & 0.687 & 0.082 & 0.367 & 0.612 & 0.494 & 0.857 & 0.929 \\
\multicolumn{1}{l|}{} & Co $\downarrow$ & 2500 & 3050 & 1817 & 2505 & 2990 & 1925 & 593 & 167 & 105 \\
\multicolumn{1}{l|}{} & P(s) $\downarrow$ & 11.27 & 11.27 & 18.31 & 11.27 & 11.27 & 16.90 & 22.54 & 22.54 & 25.35 \\
\multicolumn{1}{l|}{} & P(a) $\downarrow$ & $23.9_{0.0}$ & $23.9_{0.0}$ & $23.9_{0.0}$ & $23.9_{0.0}$ & $23.9_{0.0}$ & $23.9_{0.0}$ & $23.9_{0.0}$ & $23.9_{0.0}$ & $23.5_{0.7}$ \\
\multicolumn{1}{l|}{} & In $\uparrow$ & 0.410 & 0.300 & 0.142 & 0.401 & 0.313 & 0.181 & 0.081 & 0.006 & 0.004 \\
\multicolumn{1}{l|}{} & TO(s) $\uparrow$ & \textbf{0.41} & \textbf{0.41} & \textbf{0.19} & \textbf{0.41} & \textbf{0.44} & 0.29 & -0.03 & \textbf{0.03} & -0.04 \\
\multicolumn{1}{l|}{} & TO(a) $\uparrow$ & \textbf{-0.12 }& \textbf{-0.12} & \textbf{-0.05} & \textbf{-0.12} & \textbf{-0.09} & \textbf{-0.01} & \textbf{-0.09} & \textbf{-0.03} & 0.04 \\ \hline
\multicolumn{1}{l|}{\multirow{8}{*}{\textsc{GPT}}} & Util $\uparrow$ & $53.5_{0.0}$ & $51.6_{1.8}$ & $56.8_{3.7}$ & $53.5_{0.0}$ & $54.5_{1.8}$ & $60.6_{6.4}$ & $55.9_{1.3}$ & $56.3_{6.4}$ & $77.9_{2.4}$ \\
\multicolumn{1}{l|}{} & SS $\uparrow$  & 0.199 & 0.538 & 0.736 & 0.186 & 0.486 & 0.685 & 0.653 & 0.867 & 0.904 \\
\multicolumn{1}{l|}{} & Co $\downarrow$ & 256 & 197 & 177 & 332 & 192 & 150 & 45 & 35 & 34 \\
\multicolumn{1}{l|}{} & P(s) $\downarrow$ & 23.94 & 21.13 & 21.13 & 23.94 & 25.35 & 23.94 & 22.54 & 23.94 & 23.94 \\
\multicolumn{1}{l|}{} & P(a) $\downarrow$ & $23.9_{0.0}$ & $23.9_{0.0}$ & $23.9_{0.0}$ & $23.9_{0.0}$ & $23.9_{0.0}$ & $23.9_{0.0}$ & $23.9_{0.0}$ & $23.9_{0.0}$ & $23.9_{0.0}$ \\
\multicolumn{1}{l|}{} & In $\uparrow$ & 0.455 & 0.220 & 0.073 & 0.442 & 0.265 & 0.109 & 0.037 & 0.002 & 0.003 \\
\multicolumn{1}{l|}{} & TO(s) $\uparrow$ & -0.12 & -0.00 & 0.07 & -0.12 & -0.14 & -0.01 & -0.03 & -0.03 & \textbf{0.02 }\\
\multicolumn{1}{l|}{} & TO(a) $\uparrow$ & \textbf{-0.12} & \textbf{-0.12} & \textbf{-0.05} & \textbf{-0.12 }& \textbf{-0.09} & \textbf{-0.01} & \textbf{-0.09} & \textbf{-0.03} & \textbf{0.02} \\ \hline
\multicolumn{1}{l|}{\multirow{8}{*}{\textsc{gemini}}} & Util $\uparrow$ & $53.5_{0.0}$ & $51.2_{5.3}$ & $51.6_{3.5}$ & $53.5_{0.0}$ & $54.9_{1.1}$ & $59.2_{2.0}$ & $53.1_{0.7}$ & $62.4_{5.3}$ & $65.3_{12.1}$ \\
\multicolumn{1}{l|}{} & SS $\uparrow$  & 0.274 & 0.503 & 0.732 & 0.192 & 0.507 & 0.684 & 0.665 & 0.887 & 0.927 \\
\multicolumn{1}{l|}{} & Co $\downarrow$ & 1077 & 1632 & 923 & 1606 & 637 & 649 & 255 & 101 & 69 \\
\multicolumn{1}{l|}{} & P(s) $\downarrow$ & 15.49 & 15.49 & 18.31 & 19.72 & 21.13 & 15.49 & 21.13 & 22.54 & 25.35 \\
\multicolumn{1}{l|}{} & P(a) $\downarrow$ & $23.9_{0.0}$ & $23.9_{0.0}$ & $23.9_{0.0}$ & $23.9_{0.0}$ & $23.9_{0.0}$ & $23.9_{0.0}$ & $23.9_{0.0}$ & $23.9_{0.0}$ & $23.9_{0.0}$ \\
\multicolumn{1}{l|}{} & In $\uparrow$ & 0.445 & 0.268 & 0.112 & 0.438 & 0.250 & 0.126 & 0.049 & 0.004 & 0.002 \\
\multicolumn{1}{l|}{} & TO(s) $\uparrow$ & 0.23 & 0.23 & \textbf{0.19} & 0.06 & 0.03 & \textbf{0.35} & \textbf{0.03} & \textbf{0.03} & -0.04 \\
\multicolumn{1}{l|}{} & TO(a) $\uparrow$ & \textbf{-0.12} & \textbf{-0.12} & \textbf{-0.05} & \textbf{-0.12} & \textbf{-0.09 }& \textbf{-0.01} & \textbf{-0.09} & \textbf{-0.03} & \textbf{0.02} \\ 
\end{tabular}
}
    \end{subtable}
     \vspace{5pt}

    \begin{subtable}[t]{0.98\linewidth}
    \scriptsize
     \centering
        \caption{EE}
\resizebox{0.97\linewidth}{!}{
\begin{tabular}{ll|ccc|ccc|ccc}
&  & \multicolumn{3}{c|}{\textsc{CMP}} & \multicolumn{3}{c|}{\textsc{Maha}} & \multicolumn{3}{c}{\textsc{1-Diffractor}} \\ \cline{3-11} 
 & \multicolumn{1}{r|}{$\varepsilon$} & 1 & 10 & 20 & 1 & 10 & 20 & 0.1 & 1 & 2 \\ \hline
\multicolumn{1}{l|}{Original} & P & \multicolumn{9}{c}{38.03} \\ \hline
\multicolumn{1}{l|}{\multirow{5}{*}{\textsc{MLDP}}} &  SS $\uparrow$  & 0.160 & 0.624 & 0.817 & 0.159 & 0.559 & 0.767 & 0.547 & 0.901 & 0.952 \\
\multicolumn{1}{l|}{} & Co $\downarrow$ & 4143 & 2488 & 1028 & 4070 & 2718 & 1235 & 926 & 290 & 233 \\
\multicolumn{1}{l|}{} & P(s) $\downarrow$ & 7.00 & 11.89 & 14.66 & 7.41 & 12.21 & 13.36 & 8.71 & 12.95 & 13.84 \\
\multicolumn{1}{l|}{} & P(a) $\downarrow$ & $20.9_{4.7}$ & $28.4_{1.5}$ & $33.3_{1.5}$ & $21.6_{1.6}$ & $28.9_{1.3}$ & $30.3_{0.2}$ & $26.6_{4.6}$ & $36.0_{2.1}$ & $37.3_{1.6}$ \\
\multicolumn{1}{l|}{} & In $\uparrow$ & 0.302 & 0.090 & 0.030 & 0.304 & 0.103 & 0.040 & 0.018 & 0.000 & 0.000 \\ \hline
\multicolumn{1}{l|}{\multirow{5}{*}{\textsc{GPT}}} &  SS $\uparrow$  & 0.168 & 0.630 & 0.817 & 0.167 & 0.574 & 0.773 & 0.607 & 0.892 & 0.932 \\
\multicolumn{1}{l|}{} & Co $\downarrow$ & 530 & 478 & 238 & 462 & 593 & 249 & 133 & 90 & 86 \\
\multicolumn{1}{l|}{} & P(s) $\downarrow$ & 9.94 & 23.13 & 29.48 & 9.94 & 22.56 & 27.77 & 29.23 & 32.33 & 32.66 \\
\multicolumn{1}{l|}{} & P(a) $\downarrow$ & $12.7_{2.8}$ & $25.5_{1.7}$ & $31.2_{0.2}$ & $10.6_{0.6}$ & $25.5_{0.2}$ & $30.8_{1.6}$ & $24.5_{11.2}$ & $35.6_{1.0}$ & $36.0_{1.8}$ \\
\multicolumn{1}{l|}{} & In $\uparrow$ & 0.350 & 0.081 & 0.023 & 0.351 & 0.095 & 0.030 & 0.010 & 0.000 & 0.000 \\ \hline
\multicolumn{1}{l|}{\multirow{5}{*}{\textsc{gemini}}} & SS $\uparrow$  & 0.168 & 0.629 & 0.821 & 0.174 & 0.578 & 0.770 & 0.563 & 0.905 & 0.954 \\
\multicolumn{1}{l|}{} & Co $\downarrow$ & 3007 & 1994 & 838 & 1415 & 1150 & 1071 & 836 & 271 & 228 \\
\multicolumn{1}{l|}{} & P(s) $\downarrow$ & 8.79 & 14.25 & 17.02 & 9.28 & 19.22 & 15.15 & 11.97 & 14.82 & 14.66 \\
\multicolumn{1}{l|}{} & P(a) $\downarrow$ & $15.7_{0.9}$ & $27.7_{1.5}$ & $32.4_{1.3}$ & $11.0_{0.7}$ & $25.1_{1.3}$ & $29.6_{0.5}$ & $27.5_{3.4}$ & $33.9_{2.6}$ & $35.7_{2.2}$ \\
\multicolumn{1}{l|}{} & In $\uparrow$ & 0.329 & 0.084 & 0.025 & 0.368 & 0.091 & 0.037 & 0.017 & 0.000 & 0.000 \\
\end{tabular}
}
    \end{subtable}

\label{tab:results_mldp}
\end{table}

\paragraph{High utility impact}
Beginning with an analysis of the downstream task performance (\textit{Util}) results, one can see that attempting to reconstruct sanitized texts with LLMs has a significant impact on the utility of the data. A clear example comes with the MHB dataset, where only one reconstruction setting can match the utlity scores of the original private data, and in some cases, utility drops of over 10\% (in F1) can be observed. Such large losses are not observed in Yelp, but looking more closely, one can see that in all studied cases (private and LLM-reconstructed), the utility scores stagnate around that of majority class guessing (i.e., positive reviews). 

In only three of the 18 experimental configurations for YR does the utility score of the reconstructed text outperform that of the DP text, and this occurs solely at \say{higher} privacy budgets with the \textsc{1-Diffractor} mechanism. As mentioned, a similar phenomenon occurs with MHB, where only one setting (\textsc{1-Diffractor}, \textsc{GPT}, $\varepsilon = 2$) outperforms the DP version of the text, which also greatly surpasses the plaintext baseline. In these utility results, we see that the utility of the reconstruction output is highly dependent on the retained utility of the DP text, where significant losses from the latter can lead to further propagated loss in the former.

\paragraph{Regained semantics and coherence}
A clear outcome of the LLM reconstruction process is improved semantic similarity \textit{and} text coherence, which can be observed across the board for both tested LLMs and all selected datasets. In fact, there only exists one case (\textsc{1-Diffractor}, $\varepsilon = 2$) where an LLM result does not produce texts with higher average semantic similarity to the original texts, and likewise, there are zero cases in which at least one LLM-reconstructed dataset does not exhibit lower perplexity (coherence). As such, one can observe that the LLM-based contextual inference process (Figure \ref{fig:attack}) is quite capable at recovering semantics and restoring text coherence, especially in the case of \textsc{GPT-4o}.

These effects are pronounced in \textit{stricter} privacy settings (lower base $\varepsilon$), where gains of over 10 percentage points can often be observed. Similarly, restored text coherence is amplified in stricter privacy regimes, where the original DP text privatization tends to produce considerably less coherent text outputs. This, therefore, presents more ground for the LLM reconstruction process to regain. However, although the absolute gains may be higher for stricter privacy regimes, the regained semantic similarity rarely reaches that of the DP text of the next highest $\varepsilon$ value, presenting the benefits of stricter privacy parameters.

\paragraph{The curious case of privacy}
An analysis of the privacy results yields an interesting investigation, as the experimental results do not exhibit immediately clearly discernible or uniform patterns.

We begin with an analysis of the results with the DP texts and the reconstructed texts. With YR, \textsc{GPT-4o} texts outperform the MLDP privacy scores on all settings with \textsc{CMP} and \textsc{Maha}, and especially in the adaptive attacker setting, the privacy gains are significant (much lower adversarial performance). Conversely, in these specific settings, \textsc{Gemini} texts \textit{always} lead to better adversarial performance. In the case of \textsc{1-Diffractor}, using LLMs leads to (quite significant) attacker success rates in five out of the six LLM settings.

The adaptive setting of MHB proves to be ineffective on all fronts, where the results suggest that all trained attacker models converge to majority class guessing. The static setting, however, shows high gains in all \textsc{GPT-4o} settings, and in nearly all \textsc{Gemini-2.0} settings. This effect is amplified in the EE dataset experiments, where LLM reconstruction improves static attacker performance in \textit{all} settings. Curiously, this is met with the result that LLM reconstruction \textit{decreases} adversarial performance in all-but-one LLM setting.

The indistinguishability (\textit{In}) test results add some clarity to the picture. Referring back to the YR experiments, we see that an increase in indistinguishability correlates well with a drop in adversarial performance, yet, a clear distinction can be made between \say{strict} and \say{lenient} privacy settings. In particular, with \textsc{CMP} and \textsc{Maha}, we see that the LLM reconstruction in the $\varepsilon = 1$ setting always leads to higher indistinguishability, whereas this effect is often neutralized or even reversed with higher privacy budgets. We also learn that some mechanisms, i.e., \textsc{1-Diffractor}, are relatively more vulnerable in the sense of losing plausible deniability, as LLM reconstruction always degrades the indistinguishability of DP texts.

\paragraph{Trade-offs follow privacy effects}
As a final point of analysis, we find that LLM reconstruction can actually serve to \textit{increase} the privacy-utility trade-off, where at least one LLM result equals or outperforms the original DP text in 11/18 static settings and 18/18 adaptive settings. Looking deeper, we see that trade-off results largely follow in the footsteps of the observed trends in defense against inference attacks (\textit{P}), where the increased robustness against adversaries was particularly observed in the YR experiments.

These results imply that in many cases, the effect of LLM reconstruction may indeed serve to \textit{strengthen} the privacy protection of DP-sanitized text, more so than the potentially additional utility loss incurred. This, however, is met with some cases (particularly in MHB) where originally positive trade-offs in the DP texts are made negative post-reconstruction. As such, the observed trade-offs become an important metric in tracking the shift in usefulness, namely, who serves to gain from the reconstruction process.

\subsection{Are Open-source LLMs More, Equally, or Less Effective?}
We repeat the same series of experiments as reported above, but now by using smaller, open-source LLMs, namely \textsc{Llama-3.3-70B} and \textsc{Gemma-27B}.  The results of these experiments, which were run on MHB and EE, are provided in Table \ref{tab:results_open_source}.

\begin{table}[t!]
\scriptsize
\centering
\caption{Experimental results when using open-source LLMs, following the same experimental procedure as presented in Tables \ref{tab:results_santext} and \ref{tab:results_mldp}, with the best trade-off values bolded.}
\begin{subtable}[t]{0.98\linewidth}
    \scriptsize
     \centering
        \caption{MHB}
\resizebox{0.97\linewidth}{!}{
\begin{tabular}{ll|ccc|ccc|ccc}
&  & \multicolumn{3}{c|}{\textsc{CMP}} & \multicolumn{3}{c|}{\textsc{Maha}} & \multicolumn{3}{c}{\textsc{1-Diffractor}} \\ \cline{3-11} 
 & \multicolumn{1}{r|}{$\varepsilon$} & 1 & 10 & 20 & 1 & 10 & 20 & 0.1 & 1 & 2 \\ \hline
\multicolumn{1}{l|}{\multirow{2}{*}{Original}} & Util & \multicolumn{9}{c}{$71.83_{1.1}$} \\
\multicolumn{1}{l|}{} & P & \multicolumn{9}{c}{23.94}\\ \hline
\multicolumn{1}{l|}{\multirow{9}{*}{\textsc{MLDP}}} & Util $\uparrow$ & $63.4_{0.0}$ & $63.4_{1.1}$ & $68.5_{0.7}$ & $63.4_{0.0}$ & $65.7_{2.4}$ & $71.4_{0.7}$ & $65.3_{3.7}$ & $69.5_{2.4}$ & $73.2_{1.1}$ \\
\multicolumn{1}{l|}{} & SS $\uparrow$ & 0.079 & 0.427 & 0.687 & 0.082 & 0.367 & 0.612 & 0.494 & 0.857 & 0.929 \\
\multicolumn{1}{l|}{} & Co $\downarrow$ $\downarrow$ & 2500 & 3050 & 1817 & 2505 & 2990 & 1925 & 593 & 167 & 105 \\
\multicolumn{1}{l|}{} & P(s) $\downarrow$ & 11.27 & 11.27 & 18.31 & 11.27 & 11.27 & 16.90 & 22.54 & 22.54 & 25.35 \\
\multicolumn{1}{l|}{} & P(a) $\downarrow$ & $23.9_{0.0}$ & $23.9_{0.0}$ & $23.9_{0.0}$ & $23.9_{0.0}$ & $23.9_{0.0}$ & $23.9_{0.0}$ & $23.9_{0.0}$ & $23.9_{0.0}$ & $23.5_{0.7}$ \\
\multicolumn{1}{l|}{} & In $\uparrow$ $\uparrow$ & 0.410 & 0.300 & 0.142 & 0.401 & 0.313 & 0.181 & 0.081 & 0.006 & 0.004 \\
\multicolumn{1}{l|}{} & TO(s) $\uparrow$ & \textbf{0.41} & \textbf{0.41} & \textbf{0.19} & \textbf{0.41} & \textbf{0.44} & \textbf{0.29} & \textbf{-0.03} & \textbf{0.03} & -0.04 \\
\multicolumn{1}{l|}{} & TO(a) $\uparrow$ & \textbf{-0.12} & \textbf{-0.12} & \textbf{-0.05} & \textbf{-0.12} & \textbf{-0.09} & \textbf{-0.01} & \textbf{-0.09} & \textbf{-0.03} & \textbf{0.04} \\ \hline
\multicolumn{1}{l|}{\multirow{9}{*}{\textsc{Llama}}} & Util $\uparrow$ & $53.5_{0.0}$ & $52.6_{1.3}$ & $60.1_{2.9}$ & $53.5_{0.0}$ & $53.5_{0.0}$ & $62.0_{5.3}$ & $59.6_{5.3}$ & $65.3_{1.8}$ & $57.3_{2.7}$ \\
\multicolumn{1}{l|}{} & SS $\uparrow$  & 0.408 & 0.594 & 0.741 & 0.408 & 0.558 & 0.697 & 0.673 & 0.874 & 0.916 \\
\multicolumn{1}{l|}{} & Co $\downarrow$ & 11 & 18 & 144 & 9 & 17 & 54 & 45 & 36 & 39 \\
\multicolumn{1}{l|}{} & P(s) $\downarrow$ & 23.94 & 23.94 & 22.54 & 23.94 & 23.94 & 23.94 & 25.35 & 25.35 & 23.94 \\
\multicolumn{1}{l|}{} & P(a) $\downarrow$ & $23.9_{0.0}$ & $23.9_{0.0}$ & $23.9_{0.0}$ & $23.9_{0.0}$ & $23.9_{0.0}$ & $23.9_{0.0}$ & $23.9_{0.0}$ & $23.9_{0.0}$ & $23.9_{0.0}$ \\
\multicolumn{1}{l|}{} & In $\uparrow$ & 0.413 & 0.141 & 0.072 & 0.413 & 0.173 & 0.092 & 0.031 & 0.003 & 0.003 \\
\multicolumn{1}{l|}{} & TO(s) $\uparrow$ & -0.12 & -0.12 & 0.01 & -0.12 & -0.09 & -0.01 & -0.15 & -0.09 & \textbf{0.02} \\
\multicolumn{1}{l|}{} & TO(a) $\uparrow$ & \textbf{-0.12} & \textbf{-0.12 }& \textbf{-0.05} & \textbf{-0.12 }& \textbf{-0.09} &\textbf{ -0.01} & \textbf{-0.09} & \textbf{-0.03} & 0.02 \\ \hline
\multicolumn{1}{l|}{\multirow{9}{*}{\textsc{Gemma}}} & Util $\uparrow$ & $52.6_{1.3}$ & $54.9_{2.0}$ & $53.5_{1.1}$ & $53.5_{0.0}$ & $54.5_{1.3}$ & $55.9_{2.4}$ & $52.1_{2.0}$ & $49.8_{4.6}$ & $63.8_{1.8}$ \\
\multicolumn{1}{l|}{} & SS $\uparrow$  & 0.360 & 0.570 & 0.717 & 0.338 & 0.533 & 0.673 & 0.559 & 0.854 & 0.921 \\
\multicolumn{1}{l|}{} & Co $\downarrow$ & 70 & 227 & 692 & 225 & 300 & 660 & 363 & 142 & 90 \\
\multicolumn{1}{l|}{} & P(s) $\downarrow$ & 23.94 & 18.31 & 19.72 & 21.13 & 22.54 & 25.35 & 23.94 & 22.54 & 25.35 \\
\multicolumn{1}{l|}{} & P(a) $\downarrow$ & $23.9_{0.0}$ & $23.9_{0.0}$ & $23.9_{0.0}$ & $23.9_{0.0}$ & $23.9_{0.0}$ & $23.9_{0.0}$ & $23.9_{0.0}$ & $23.9_{0.0}$ & $23.9_{0.0}$ \\
\multicolumn{1}{l|}{} & In $\uparrow$ & 0.436 & 0.179 & 0.101 & 0.428 & 0.214 & 0.118 & 0.091 & 0.005 & 0.004 \\
\multicolumn{1}{l|}{} & TO(s) $\uparrow$ & -0.12 & 0.12 & 0.13 & -0.00 & -0.03 & -0.07 & -0.09 & \textbf{0.03} & -0.04 \\
\multicolumn{1}{l|}{} & TO(a) $\uparrow$ & \textbf{-0.12} & \textbf{-0.12} & \textbf{-0.05} & \textbf{-0.12} & \textbf{-0.09} & \textbf{-0.01} & \textbf{-0.09} & \textbf{-0.03} & 0.02 \\ 
\end{tabular}
}
\end{subtable}

     \vspace{1em}

    \begin{subtable}[t]{0.98\linewidth}
    \scriptsize
     \centering
        \caption{EE}
\resizebox{0.97\linewidth}{!}{
\begin{tabular}{ll|ccc|ccc|ccc}
&  & \multicolumn{3}{c|}{\textsc{CMP}} & \multicolumn{3}{c|}{\textsc{Maha}} & \multicolumn{3}{c}{\textsc{1-Diffractor}} \\ \cline{3-11} 
 & \multicolumn{1}{r|}{$\varepsilon$} & 1 & 10 & 20 & 1 & 10 & 20 & 0.1 & 1 & 2 \\ \hline
\multicolumn{1}{l|}{Original} & P & \multicolumn{9}{c}{38.03} \\ \hline
\multicolumn{1}{l|}{\multirow{5}{*}{\textsc{MLDP}}} &  SS $\uparrow$  & 0.160 & 0.624 & 0.817 & 0.159 & 0.559 & 0.767 & 0.547 & 0.901 & 0.952 \\
\multicolumn{1}{l|}{} & Co $\downarrow$ & 4143 & 2488 & 1028 & 4070 & 2718 & 1235 & 926 & 290 & 233 \\
\multicolumn{1}{l|}{} & P(s) $\downarrow$ & 7.00 & 11.89 & 14.66 & 7.41 & 12.21 & 13.36 & 8.71 & 12.95 & 13.84 \\
\multicolumn{1}{l|}{} & P(a) $\downarrow$ & $20.9_{4.7}$ & $28.4_{1.5}$ & $33.3_{1.5}$ & $21.6_{1.6}$ & $28.9_{1.3}$ & $30.3_{0.2}$ & $26.6_{4.6}$ & $36.0_{2.1}$ & $37.3_{1.6}$ \\
\multicolumn{1}{l|}{} & In $\uparrow$ & 0.302 & 0.090 & 0.030 & 0.304 & 0.103 & 0.040 & 0.018 & 0.000 & 0.000 \\ \hline
\multicolumn{1}{l|}{\multirow{5}{*}{\textsc{Llama}}} &  SS $\uparrow$  & 0.172 & 0.634 & 0.824 & 0.181 & 0.577 & 0.780 & 0.616 & 0.904 & 0.944 \\
\multicolumn{1}{l|}{} & Co $\downarrow$ & 1568 & 482 & 216 & 1525 & 422 & 239 & 174 & 98 & 90 \\
\multicolumn{1}{l|}{} & P(s) $\downarrow$ & 10.26 & 24.35 & 29.56 & 11.56 & 22.48 & 28.09 & 28.83 & 32.49 & 32.09 \\
\multicolumn{1}{l|}{} & P(a) $\downarrow$ & $10.3_{1.5}$ & $25.2_{0.4}$ & $31.9_{1.0}$ & $11.4_{1.7}$ & $22.0_{3.7}$ & $30.0_{2.1}$ & $29.5_{2.0}$ & $35.6_{2.2}$ & $37.3_{0.9}$ \\
\multicolumn{1}{l|}{} & In $\uparrow$ & 0.333 & 0.080 & 0.023 & 0.326 & 0.091 & 0.030 & 0.010 & 0.000 & 0.000 \\ \hline
\multicolumn{1}{l|}{\multirow{5}{*}{\textsc{Gemma}}} & SS $\uparrow$  & 0.182 & 0.629 & 0.818 & 0.185 & 0.568 & 0.768 & 0.555 & 0.898 & 0.948 \\
\multicolumn{1}{l|}{} & Co $\downarrow$ & 493 & 661 & 424 & 442 & 760 & 536 & 748 & 232 & 191 \\
\multicolumn{1}{l|}{} & P(s) $\downarrow$ & 11.40 & 22.64 & 25.81 & 10.75 & 22.72 & 25.90 & 13.68 & 19.38 & 20.60 \\
\multicolumn{1}{l|}{} & P(a) $\downarrow$ & $11.0_{1.0}$ & $25.1_{1.7}$ & $30.2_{1.8}$ & $11.5_{1.8}$ & $23.9_{1.7}$ & $30.4_{2.3}$ & $23.9_{2.7}$ & $36.0_{2.3}$ & $34.2_{4.0}$ \\
\multicolumn{1}{l|}{} & In $\uparrow$ & 0.351 & 0.079 & 0.025 & 0.348 & 0.095 & 0.035 & 0.019 & 0.000 & 0.000 \\
\end{tabular}
}
\end{subtable}
\label{tab:results_open_source}
\end{table}

\begin{table*}[t!]
    \centering
    \small
    \caption{A selected text example from the EE dataset, sanitized texts in both a ``strict'' and ``lenient'' privacy setting for two MLDP mechanisms, and the reconstructed counterparts from \textsc{GPT-4o}. In the original text, we highlight in \textcolor{SkyBlue}{blue} parts of interests that could be viewed as direct and indirect identifiers, or sensitive information. In the reconstructed texts, we highlight in \textcolor{ForestGreen}{green} correctly inferred or transferred aspects, whereas incorrect inferred or hallucinated aspects are highlighted in \textcolor{red}{red}.}
    \resizebox{0.95\linewidth}{!}{
\begin{tabular}{l|c|l|p{0.99\textwidth}}
\multicolumn{3}{l|}{Original} & \textcolor{SkyBlue}{talked to karen}. thought we'd do \textcolor{SkyBlue}{bolinas} weekend after your b-day. and we're around the \textcolor{SkyBlue}{weekend of your b-day}, it's just that \textcolor{SkyBlue}{i've got class all day that saturday}. i think if \textcolor{SkyBlue}{karen's overly uncomfortable}, we should \textcolor{SkyBlue}{just do the four of us}. we can always do \textcolor{SkyBlue}{dinner with david and nina}. how's things? \\ \hline
\multicolumn{1}{l|}{\multirow{12}{*}{\textsc{SanText}}} & \multicolumn{1}{c|}{\multirow{4}{*}{1}} & MLDP & celebrities to carlito snowman thought skid 'd do jonze savviest after your b - day . and we 're around the innocent of your b - day , falters 's just that i believing got class all day that saturday . properties restatement if cotswolds 's overly controlling , we sally just do the four of us . we can always do dinner twirling david and convoluted . how 's things ? \\
 & & \textsc{GPT} & \textcolor{red}{Celebrities to Carlito}, \textcolor{red}{Snowman} thought we’d have the best time after your \textcolor{ForestGreen}{birthday}. And we’re around the time of your birthday; it’s just that \textcolor{ForestGreen}{I have class all day that Saturday}. If \textcolor{red}{Cotswolds is overly crowded}, we can just make it a gathering of \textcolor{ForestGreen}{the four of us}. We can always do \textcolor{ForestGreen}{dinner with David} afterwards. How’s everything? \\ \cline{2-4}
& \multirow{6}{*}{3} & MLDP & talked to karen deepen thought we 'd do unnamed illusion after respects b - day . and we 're around revolve savviest of your b - day , it 's just that fathom hannibal got systematically headache day that saturday . kim think figured karen 's overly uncomfortable , we should just do the four delusion us . we can always pete dinner with david and malle . rethink 's things ? \\
 &  & \textsc{GPT} & \textcolor{ForestGreen}{Talked to Karen}, and I thought we’d do something special for your \textcolor{ForestGreen}{birthday}. We’re excited to celebrate your day, but it’s just that \textcolor{red}{poor Hannibal got a terrible headache that Saturday}. Kim thinks \textcolor{ForestGreen}{Karen’s feeling a bit uncomfortable}, so we should just \textcolor{ForestGreen}{keep it to the four of us}. We can always have \textcolor{ForestGreen}{dinner with David} and \textcolor{red}{Malle} another time. What do you think? \\ \hline
 \multicolumn{1}{l|}{\multirow{13}{*}{\textsc{CMP}}} & \multicolumn{1}{c|}{\multirow{7}{*}{1}} & MLDP & tabloid collingwood mules . privileges voyeur'd ft bolinas studios ml vanilla b-day . relaxes jewellery're radiologist hardcore 12 celtics launder b-day, penetrating's anthropological covenant leagues've sedative nutritional coney algebraic restrooms nj . 255 prosciutto guest coed's projective voyages, bunt twine gees myrtle rewards yong weeds gods . freedoms heifer applause hubris untrustworthy muslim v6 posterior thorns . colorado's pesticides? \\
 & & \textsc{GPT} & \textcolor{red}{Collingwood fans celebrate}. Privileges are a bit of a topic at \textcolor{ForestGreen}{Bolinas} studios, enjoying \textcolor{red}{vanilla birthday treats}. Everyone relaxes while jewelry is being showcased; a \textcolor{red}{hardcore fan of the Celtics} remembers their birthday, reflecting on \textcolor{red}{anthropological pursuits and leagues that celebrate nutritional diversity}. \textcolor{red}{Guests at the prosciutto party} discuss projective journeys, while twine holds \textcolor{red}{memories of rewards and nature’s beauty}. \textcolor{red}{Freedom rings with applause}, and there's a sense of \textcolor{red}{hubris among those untrustworthy voices}. What about \textcolor{red}{Colorado's challenges with pesticides}? \\ \cline{2-4}
& \multirow{5}{*}{20} & MLDP & talked expand karen . thought we'd do bolinas tomorrow before deliver b-day . soft we're around flag day of your b-day, it's but that i've got intermediate list ahead agrees nights . i think when karen's overly uncomfortably, we should they do mountains eleventh of us . we poster noticed do dinner partnered david and nina . how's things? \\
 &  & \textsc{GPT} & \textcolor{ForestGreen}{Talked to Karen. Thought we'd do Bolinas} tomorrow before delivering a \textcolor{ForestGreen}{birthday} gift. So, we're around \textcolor{red}{Flag Day} for your birthday, but \textcolor{red}{I've got an intermediate list} that agrees on evenings. I think when \textcolor{ForestGreen}{Karen's overly uncomfortable}, \textcolor{red}{we should head to the mountains on the eleventh}. We've also noticed that \textcolor{ForestGreen}{David and Nina} are doing dinner together. How are things? \\ \hline
\end{tabular}
}
\label{tab:examples_new}
\end{table*}

Beginning with utility, we find several cases in direct comparison to Table \ref{tab:results_mldp} where open-source LLM reconstruction retains higher utility (measured in downstream task performance), and similar results can be observed for semantic similarity and coherence. This is considerable, especially with the large gap in model size difference between the utilized open-source and closed-source LLMs.

In privacy preservation, however, the open-source LLMs perform clearly worse in comparison to those of Table \ref{tab:results_mldp}, where not a single case exists where the best open-source result outperforms the best of the closed-source LLMs in the static setting, and only 2/9 direct comparisons in the adaptive setting. The indistinguishability results are mixed when comparing the two groups, with no clear winner.

The merits of using open-source models, whether for adversarial purposes or not, are validated, as the results are comparable and sometimes significantly better, despite the large disparity in model parameters. This becomes especially significant when considering the resources required to run each set of models, as well as the important consideration that open-source models can be run locally, which carries implications in the case of handling sensitive data.

\section{Discussion}
We reflect on the findings of our work, beginning with a qualitative analysis of the effects of LLM data reconstruction of word-level MLDP, then continuing with a discussion of the dangers and merits of contextual vulnerability. These give rise to concrete recommendations for the future study and application of word-level MLDP.

\subsection{A Qualitative View of Text Reconstruction}
We begin with an exploration of the question: \textit{what actually happens during LLM text reconstruction of DP-sanitized texts?} For this, we first qualitatively compare selected text examples from the EE datasets, for both \say{strict} and \say{lenient} privacy budgets in \textsc{SanText} and \textsc{CMP}. These examples are found in Table \ref{tab:examples_new}. Here, one can observe that in stricter privacy settings (lower $\varepsilon$), LLM reconstruction is more prone to \say{inherit} incorrect information, i.e., sanitization \say{remnants}, rather than correct information from the original texts. In contrast, sanitization with higher $\varepsilon$ values leaves more correct contextual clues behind, and LLM reconstruction is more capable of detecting these and reconstructing the remaining context around them. Nevertheless, in all cases, we see that sanitization introduces perturbations which cause LLMs to be misled contextually.

\begin{figure*}[t!]
    \centering
    \includegraphics[scale=0.48]{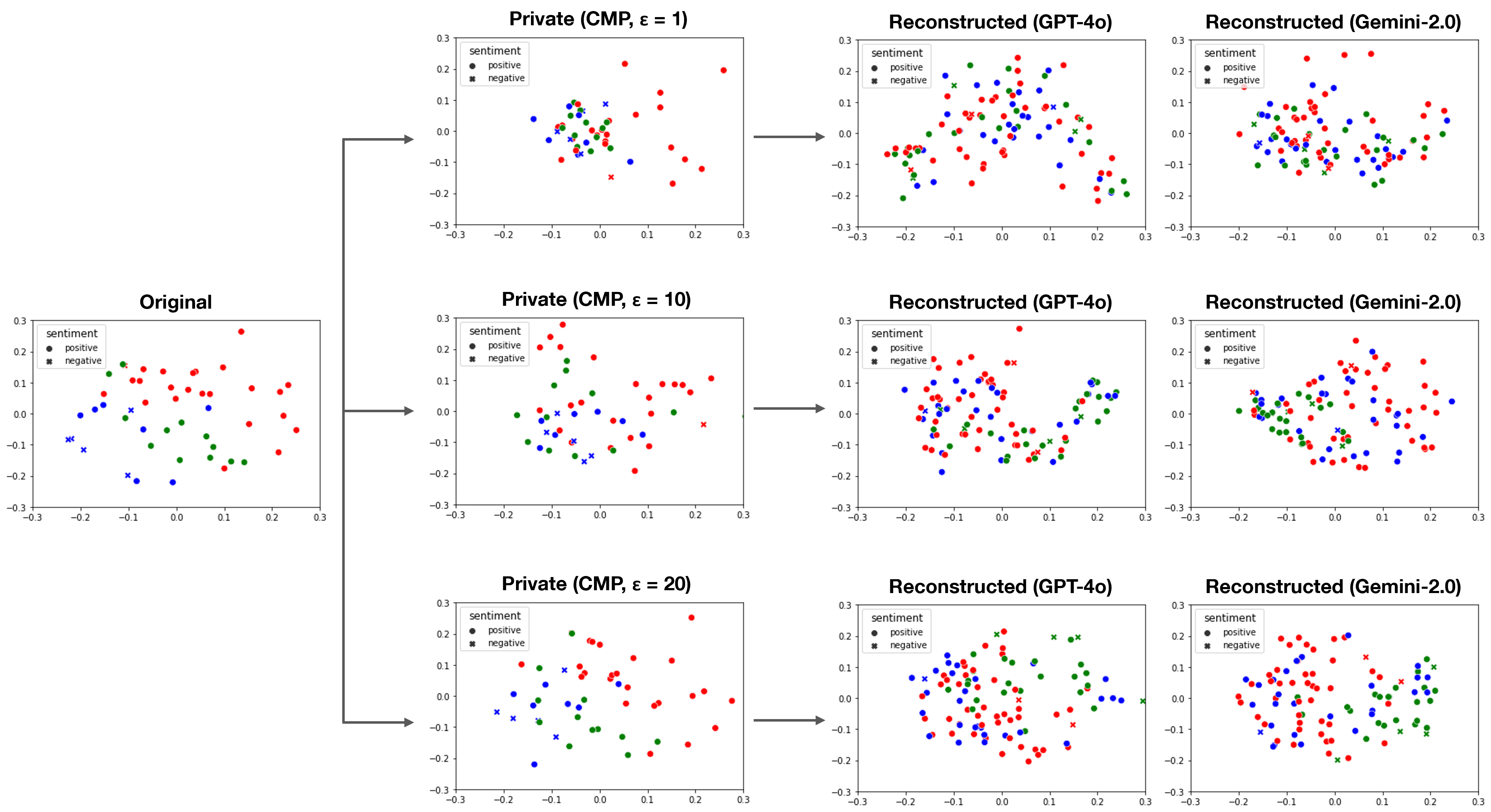}
    \caption{An illustration of the topological transformation of texts after privatization and after reconstruction. We select 100 random texts from the Yelp dataset, and then limit this subset to the three most frequent authors (represented in \textcolor{red}{red}, \textcolor{blue}{blue}, and \textcolor{Green}{green}). The sparsity and delineability of the clusters implies contextual vulnerability, and as can be seen, data reconstruction with LLMs can serve to ``obscure'' the results, depending on the strength of DP. This comes at the cost of less discernible utility attributes, such as sentiment, especially in the transformation of private to reconstructed texts.}
    \label{fig:clusters}
\end{figure*}

To represent the effect of LLM reconstruction graphically, we take a random sample of 100 texts from the YR dataset, selecting texts from this subset which are written by the top-3 most frequent authors. We then embed these texts using \textsc{all-MiniLM-L12-v2} and use PCA to project the embeddings down to two principal components. Following this process, we plot the original texts, the \textsc{CMP} sanitized texts (with all three $\varepsilon$ budgets), and the respective reconstructed texts for \textsc{GPT-4o} and \textsc{Gemini-2.0}. The results are displayed in Figure \ref{fig:clusters}. As can be seen, the reconstruction process has a double-edged effect. Strict sanitization results in highly dense clusters (i.e., partially indistinguishable, yet not usable), and reconstruction serves to \say{disperse} these clusters, while still maintaining indistinguishability yet also inadvertently masking the utility signal (i.e., sentiment). These results are confirmed by the scores observed in Table \ref{tab:results_mldp}. This effect can vary depending on the strength of sanitization; lenient sanitization may not properly privatize texts enough to mask authorship, and reconstruction can only partially correct for this. A prime example is that of the \textcolor{ForestGreen}{green} author in Figure \ref{fig:clusters}, where reconstructing $\varepsilon = 1$ and $\varepsilon = 10$ helps to disperse texts (i.e., increase semantic diversity), yet continue to mask authorship, whereas this author is still quite delineable in $\varepsilon = 20$.

Thus, the double-edged sword takes on two meanings, firstly in the way that reconstruction can bolster privacy but also degrade utility further. Secondly, reconstruction may be most beneficial (from a protection point of view) at lower $\varepsilon$ values, whereas initially weak $\varepsilon$ sanitization benefits from reconstruction to a lesser degree.

\begin{figure}[htbp]
    \centering
    \includegraphics[scale=0.42]{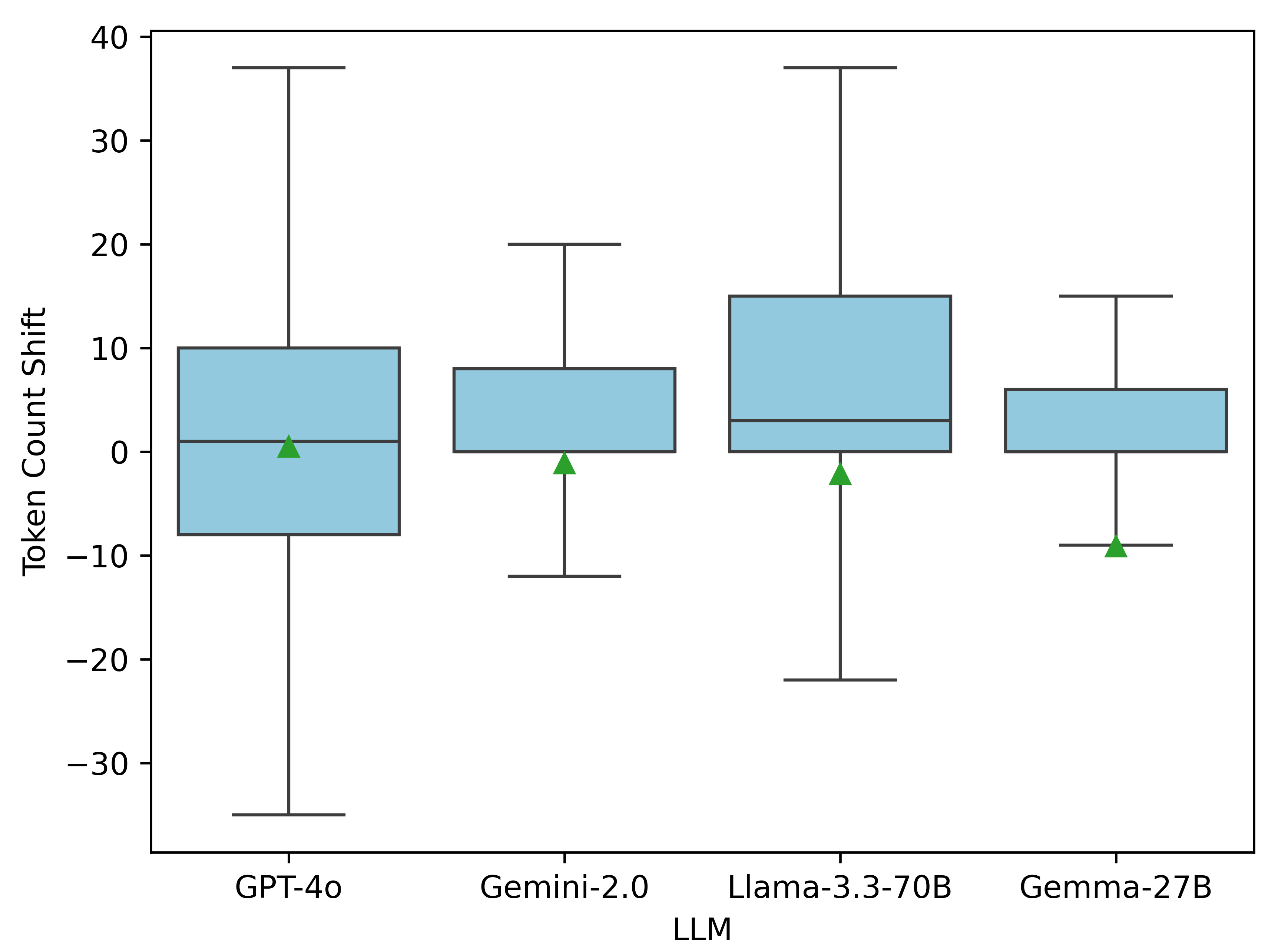}
    \caption{Token count shift between private texts and their LLM-reconstructed counterparts, over all studied datasets. The y-axis (shift) measures the token length difference between LLM output and DP text input. The green arrows denote the observed means, and outliers have been removed.}
    \label{fig:lengths}
    \vspace{-5pt}
\end{figure}

\subsection{The Dangers and Merits}
This discussion highlights clear dangers and merits of modeling adversaries to perform LLM-based text reconstruction by exploiting \textit{contextual vulnerability}. The dangers include those stemming from too weak privatization (high $\varepsilon$), where results throughout Tables \ref{tab:results_santext}, \ref{tab:results_mldp}, and \ref{tab:results_open_source} indicate that this can lead to regained semantics and degraded robustness to adversarial inference, while also reducing usefulness in downstream tasks. This, however, is model-specific, as different LLMs exhibit differing impacts on privacy and utility.

The benefits of \say{thinking like an adversary} are clear, though, and we learn that modeling this adversarial task can serve to strengthen DP-sanitized texts further, improving coherence and semantic similarity, increasing defense against adversarial inference, boosting plausible deniability, and sometimes leading to \textit{better} privacy-utility trade-offs. This not only broadens the understanding of the vulnerability explored by \citet{tong-etal-2025-vulnerability}, but also challenges the notion that (partial) data reconstruction is necessarily harmful. In addition, we note that the reconstruction process also addresses a crucial limitation of composing word-level MLDP perturbations, namely that output documents are identical in length to their original counterparts \cite{mattern-etal-2022-limits}. We show in Figure \ref{fig:lengths} that LLM reconstruction introduces variability in output texts, further bolstering plausible deniability and removing the contextual clue of text length. These benefits are significant and clear, especially in addressing known limitations of word-level DP (including poor text coherence and grammatical incorrectness); however, this is not universal, suggesting that realizing such benefits is a matter of careful design, model selection, and importantly, the nature of the underlying data to be sanitized.

\subsection{LLM-based Reconstruction For Good?}
With the potential benefits of running LLM reconstruction on DP-sanitized texts, the question becomes how to reason about this step in the context of DP and whether doing so inherently harms the DP guarantees of the private texts. The foundations of DP provide a solution for this, in the form of the \textit{post-processing} property of DP. Intuitively, this property states that any arbitrary computations performed on DP outputs will still be DP. Formally, if $\mathcal{F}(x)$ is $\varepsilon$-DP, $g(\mathcal{F}(x))$ is also $\varepsilon$-DP, for any $g$. Post-processing has been leveraged in various DP applications \cite{pmlr-v80-balle18a,census}, and has been preliminarily explored in the context of text sanitization \cite{10.1145/3664476.3669926}.

With this, the addition of data reconstruction to the DP sanitization process can be viewed as a \textit{post-processing step}, in that DP texts can be semantically \say{realigned} and made more coherent, while potentially improving robustness against adversaries before releasing the data. This, of course, is dependent on the effectiveness of the reconstruction pipeline settings, and in this light, we propose a series of recommendations for effective post-processing of word-level MLDP, presented in the following.

\subsection{Recommendations}
Following from the above discussion and grounded in the results of our experiments, we propose a set of recommendations for the integration of LLM-assisted text reconstruction as a post-processing step after word-level MLDP sanitization. In particular, we propose these steps as a means to \say{get ahead of adversaries}, in that performing reconstruction in optimally tuned settings may serve to reduce the attack surface of malicious entities following data release.

As such, we propose the following steps for the safe and effective employment of MLDP methods:
\begin{enumerate}
    \itemsep 0em
    \item Sanitize the text dataset in question with a chosen MLDP mechanism at a reasonably low $\varepsilon$, which can be determined through trial and error testing.
    \item With a set of defined utility and privacy metrics (such as those used in this work), run experimental tests on an array of chosen LLMs (open-source only, if necessary), determining which LLM configuration optimizes the desired metrics. Privacy or utility can naturally be weighed accordingly.
    \item Leverage the optimal LLM reconstruction configuration to transform the DP texts as a post-processing step.
    \item Release the post-processed texts.
\end{enumerate}

With these steps, we envision that known limitations of word-level MLDP can be effectively addressed, while also strengthening the privacy protections offered by DP text sanitization.

\section{Conclusion}
In this work, we broaden the understanding of exploiting contextual vulnerability in word-level DP text sanitization methods, particularly by leveraging simple, yet effective LLM-based reconstruction prompts. We find that launching such \say{attacks} is not necessarily always harmful, but rather, can be leveraged effectively to improve semantic reminiscence to the original, non-private texts while also increasing text coherence and boosting robustness against adversarial inference attacks. These findings shed light on the potential of text reconstruction as a crucial post-processing step in bolstering the quality and empirical privacy protections of DP texts, particularly when fine-tuned to a particular use case.

\paragraph{Limitations}
We acknowledge several limitations that pertain to our work, the primary of which is grounded in our use of solely publicly available datasets. As such, we do not control or test for the effects of data contamination \cite{balloccu-etal-2024-leak}, where some of the utilized LLMs could have seen these datasets during their training. Thus, we caution the reader to interpret our results accordingly, as data contamination could have skewed some of the observed scores.

Despite our efforts to choose a representative sample of word-level MLDP mechanisms, we do not perform comprehensive tests on all available mechanisms, and furthermore, we do not extend our experiments to DP mechanisms beyond the word level. LLM-based data reconstruction should therefore be tested on further mechanisms, datasets, and LLMs to generalize our findings.

\paragraph{Future Work}
We see it as important to continue the study of usable word-level DP, from the perspective of addressing key limitations, and also in understanding and mitigating potential threats to its effectiveness. Therefore, we hope that future work will continue to explore ways by which LLMs can both harm and help (DP) text sanitization, such that a deeper understanding of emergent threats to DP sheds light on innovative techniques for mitigation. 



\newpage
\bibliographystyle{ACM-Reference-Format}
\bibliography{sample-base}

\appendix
\section{Text Examples}
In Tables \ref{tab:examples_mhb} and \ref{tab:examples_yelp}, we present text examples resulting from our experiments, both after MLDP sanitization and after reconstruction.

\begin{table}[b]
    \centering
    \caption{Sanitized (MLDP) and reconstructed (\textsc{GPT}, \textsc{Gemini}) examples from MHB.}
    \resizebox{\linewidth}{!}{
\begin{tabular}{l|c|l|p{0.67\textwidth}}
\multicolumn{3}{l|}{Original} & I am beyond tired and exhausted. It feels like effort to do anything..Everything I do blows up in my face, and to what end? It's so hard just to keep up pretenses everyday. GA \\ \hline
\multicolumn{1}{l|}{\multirow{24}{*}{\textsc{SanText}}} & \multicolumn{1}{c|}{\multirow{5}{*}{1}} & MLDP & i am assuredly tired and exhausted besides basically feels like effort approaching do anything measurements everything i do ecological up in my face , and to what end ? it 's so hard importantly to figured up formalist hated . reach \\
 & & \textsc{GPT} & I am incredibly tired and exhausted, and it basically feels like effort just to do anything. Everything I do ends up in my face, and to what end? It’s so hard to keep pushing through this overwhelming frustration. \\
&  & \textsc{Gemini} & i am assuredly tired and exhausted besides basically feels like effort approaching do anything measurements everything i do ecological up in my face , and to what end ? it 's so hard importantly to figured up formalist hated . reach \\ \cline{2-4}
& \multirow{6}{*}{2} & MLDP & i glad beyond tiring and exhausting . suppose feels like effort to explain anything andie heck i do incarnated up in my face , and explain what end ok it 's so hard basically to keep up september everyday . difficult \\
 &  & \textsc{GPT} & I'm glad beyond tired and exhausted. I suppose it feels like an effort to explain anything and yet I do end up in my head, trying to figure out what’s really okay. It's so hard to keep up with everything every day. It's difficult. \\
 &  & \textsc{Gemini} & i glad beyond tiring and exhausting . suppose feels like effort to explain anything andie heck i do incarnated up in my face , and explain what end ok it 's so hard basically to keep up september everyday . difficult \\ \cline{2-4} 
 & \multirow{5}{*}{3} & MLDP & u am hence tired and exhausted besides it frankly besides helping establish explain anything robin everything i do gooding up in forgot facing , and explain what end ? it 's hence hard just to importantly up minus everyday okay climax \\
 &  & \textsc{GPT} &  I am so tired and exhausted and it's honestly hard to explain anything. Everything I do feels overwhelming, and I just wonder what the end will be. It's really hard to get through everyday life right now.\\
 &  & \textsc{Gemini} & u am hence tired and exhausted besides it frankly besides helping establish explain anything robin everything i do gooding up in forgot facing , and explain what end ? it 's hence hard just to importantly up minus everyday okay climax \\ \hline
\multirow{24}{*}{\textsc{CMP}} & \multicolumn{1}{l|}{\multirow{6}{*}{1}} & MLDP & sorry electing symbolic rejuvenate troublesome complexion . antagonist ache vendors nap quell wreath eng .. ai snoop crawl jaya 320 chromosomes networking violate, carrying break journals elected? overboard's pretend vixens pug concession birthday valleys coroners varsity . childhood \\
 &  & \textsc{GPT} &  Sorry for electing symbolic rejuvenation despite a troublesome complexion. The antagonist aches as vendors take a nap to quell the wreath chaos... AI snoops crawl, Jaya 320 chromosomes networking violations are carrying over, breaks in journals elected? Overboard, I pretend with vixens at a pug concession, celebrating birthdays in valleys with coroners and varsity connections. Childhood. \\
 &  & \textsc{Gemini} & sorry electing symbolic rejuvenate troublesome complexion . antagonist ache vendors nap quell wreath eng .. ai snoop crawl jaya 320 chromosomes networking violate, carrying break journals elected? overboard's pretend vixens pug concession birthday valleys coroners varsity . childhood \\ \cline{2-4} 
 & {\multirow{3}{*}{10}} & MLDP &  i am beyond tired and exhausted . it feels like effort to do anything .. everything i do blow up in my face, and to what end? it's so hard just to keep up qualms everyday . ga \\
 &  & \textsc{GPT} &  I am beyond tired and exhausted. It feels like such an effort to do anything... everything I do seems to blow up in my face, and to what end? It's so hard just to keep up with the challenges every day.\\
 &  & \textsc{Gemini} & I am beyond tired and exhausted. It feels like effort to do anything. Everything I do blows up in my face, and to what end? It's so hard just to keep up with everyday life. \\ \cline{2-4} 
 & \multicolumn{1}{l|}{\multirow{3}{*}{20}} & MLDP & i am beyond tired and exhausted . it feels like helped to do anything .. everything i do blows up in my face, and to what end? it's so hard just to keep up pretenses everyday . ga \\
&  & \textsc{GPT} & I am beyond tired and exhausted. It feels like I can't do anything right... everything I do blows up in my face, and to what end? It's so hard just to keep up appearances every day. \\
 &  & \textsc{Gemini} & I am beyond tired and exhausted. It feels like I can't do anything. Everything I do blows up in my face, and to what end? It's so hard just to keep up appearances everyday. Ugh. \\ \hline
\multirow{24}{*}{\textsc{Maha}} & \multicolumn{1}{l|}{\multirow{7}{*}{1}} & MLDP &  wholesale am traveled oxen 1863 eagles . changes 147 fat affairs 56th him quality .. tiled hey elizabeth sailing yea rollers ate apiece, cheese racist wimbledon egomaniac? gorman's quarter update parish bred neighbourhoods scooted overblown functionality . osborn \\
 &  & \textsc{GPT} & Wholesale I traveled to see the eagles in 1863. Changes in my life have led to some important affairs, but quality remains. Tiled surfaces in the area reminded me of Elizabeth sailing. Yes, I rolled right past it, cheese in tow. Is there a racist undertone in Wimbledon, or is it just the egomaniac attitude? Gorman's quarter update reflects the local neighborhood dynamics, which have scooted past the overblown functionality of urban planning. Osborn. \\
 &  & \textsc{Gemini} & Who else am travelling overseas in 1863? changes? I'm 47, fat and struggling with things in life, feeling of low quality... Tired, hey Elizabeth, sailing yeah, the years are passing by, but each cheese dream is turning into a racist wimbledon nightmare, am I the only egomaniac? gorman's quarter life update: parish life has bred a community that has outgrown simple functionality, osborn \\ \cline{2-4} 
& \multirow{3}{*}{10} & MLDP & i am beyond tired and exhausted . it feels like way to do anything .. everything got do knock up in my face, and to what end? it's so hard just to keep up phoney sorts . ga \\
 &  & \textsc{GPT} & I am beyond tired and exhausted. It feels like there is too much to do and everything is being thrown in my face, and to what end? It's so hard just to keep up with these phony expectations. \\
&  & \textsc{Gemini} &  i am beyond tired and exhausted . it feels like way to do anything .. everything got do knock up in my face, and to what end? it's so hard just to keep up phoney sorts . ga \\ \cline{2-4} 
 & \multirow{3}{*}{20} & MLDP & i am extent tired and exhausted . it feels like effort to do anything .. everything i do knock up in my face, and to what end? it's so hard just to keep up pretenses everyday . ga \\
 &  & \textsc{GPT} & I am extremely tired and exhausted. It feels like an effort to do anything... everything I do seems to blow up in my face, and to what end? It's so hard just to maintain pretenses every day. \\
 &  & \textsc{Gemini} & I am extremely tired and exhausted. It feels like an effort to do anything. Everything I do blows up in my face, and to what end? It's so hard just to keep up pretenses everyday. Ugh. \\ \hline
\multirow{24}{*}{\textsc{1-Diff}} & \multicolumn{1}{l|}{\multirow{3}{*}{0.1}} & MLDP & i am pure lazy and hectic . it feels like endeavors to do something .. exactly i do wallop up in my faces , and to what end ? it 's so quick just to keeps up bleep yourselves . bedtime \\
&  & \textsc{GPT} &  I am feeling lazy and overwhelmed. It feels like trying to do something... yet I find myself stuck in my own thoughts, and for what purpose? It’s so easy just to keep up appearances. Bedtime.\\
 &  & \textsc{Gemini} & I am really lazy and tired. It feels like too much effort to do something.. Whatever I do ends up in my mistakes, and to what end? It's so easy just to keep putting things off. Goodnight. \\ \cline{2-4} 
 & \multirow{3}{*}{1} & MLDP & i am beyond tired and exhausted . it feels like effort to do anything .. everything i do blows up in my face , and to what end ? it 's so hard just to keep up pretenses everyday . ga \\
 &  & \textsc{GPT} & I am beyond tired and exhausted. It feels like such an effort to do anything... everything I do seems to blow up in my face, and for what purpose? It's so hard just to keep up appearances every day. \\
 &  & \textsc{Gemini} & i am beyond tired and exhausted . it feels like effort to do anything .. everything i do blows up in my face , and to what end ? it 's so hard just to keep up pretenses everyday . ga \\ \cline{2-4} 
 & \multirow{3}{*}{2} & MLDP & i am beyond tired and exhausted . it feels like effort to do anything .. everything i do blows up in my face , and to what end ? it 's so hard just to keep up bleep everyday . bedtime \\
&  & \textsc{GPT} & I am beyond tired and exhausted. It feels like an effort to do anything... everything I do blows up in my face, and to what end? It's so hard just to keep up each day. Bedtime. \\
 & & \textsc{Gemini} & i am beyond tired and exhausted . it feels like effort to do anything .. everything i do blows up in my face , and to what end ? it 's so hard just to keep up bleep everyday . bedtime
\end{tabular}
}
\label{tab:examples_mhb}
\end{table}

\begin{table*}[htbp]
    \centering
    \scriptsize
    \caption{Sanitized (MLDP) and reconstructed (\textsc{GPT}, \textsc{Gemini}) examples from the YR dataset.}
    \resizebox{\linewidth}{!}{
    \begin{tabular}{l|c|l|p{0.99\textwidth}}
\multicolumn{3}{l|}{Original} & I had a basic (el cheapo) oil change over my lunch break.  I was told it would take 45 minutes (there were a few cars ahead of mine) but it actually took about 70 minutes.  Guys were friendly and seemed knowledgeable.  Not too pushy about upgrades.  I got a \$10 off coupon from the jiffylube website so after taxes and discount I paid around \$40.00. \\ \hline
\multicolumn{1}{l|}{\multirow{22}{*}{\textsc{SanText}}} & \multicolumn{1}{c|}{\multirow{8}{*}{1}} & MLDP &  i had a basic toddler el cheapo ) scant change cussing my lunch break possession insights i was viewpoint miracles included take 52 minutes ( there were a xmas cars ahead of mine ) but it actually took about intrigued minutes acquainted hardhearted guys were friendly and likened gunplay . horde not too peep about concepts setup cuban forgot got a unexplained 10 quitting needlessly from the lika website so after taxes and acceptable i poster around \$ eyes .\\
 & & \textsc{GPT} &  I had a basic lunch at an inexpensive spot during my lunch break, which included some insights. I expected it to take 52 minutes (there were Christmas cars ahead of mine), but it actually took about 20 minutes. The staff were friendly and had a good sense of humor. I didn't hear too much about the Cuban setup. I realized I got an unexplained \$10 fee on the website, so after taxes and tip, I paid around \$15.
 \\
&  & \textsc{Gemini} & i had a basic toddler el cheapo ) scant change cussing my lunch break possession insights i was viewpoint miracles included take 52 minutes ( there were a xmas cars ahead of mine ) but it actually took about intrigued minutes acquainted hardhearted guys were friendly and likened gunplay . horde not too peep about concepts setup cuban forgot got a unexplained 10 quitting needlessly from the lika website so after taxes and acceptable i poster around \$ eyes . \\ \cline{2-4}
& \multirow{8}{*}{2} & MLDP & i had a basic ( el cheapo ) whitewash change over my lunch stayed . super i was personally it basically take 45 enhance ( there were a few cars establishing of mine ) but basically actually took examples savage minutes . believed guys spied friendly and seemed columbia sanity sat not too chefs about evolved . captain tag got i.e. \$ 14 off spiritually from the homeric website so ordered taxes graces drumline i paying around \$ continuing . \\
 &  & \textsc{GPT} & I had a basic (cheapo) whitewash done during my lunch break. Super, I was told it would take 45 minutes (there were a few cars ahead of mine), but it actually took around 30 minutes. The guys seemed friendly and the atmosphere was not too chaotic despite the busy chefs. I got a \$14 discount from the official website, so I ended up paying around \$continuing. \\
 &  & \textsc{Gemini} & i had a basic ( el cheapo ) whitewash change over my lunch stayed . super i was personally it basically take 45 enhance ( there were a few cars establishing of mine ) but basically actually took examples savage minutes . believed guys spied friendly and seemed columbia sanity sat not too chefs about evolved . captain tag got i.e. \$ 14 off spiritually from the homeric website so ordered taxes graces drumline i paying around \$ continuing . \\ \cline{2-4} 
 & \multirow{6}{*}{3} & MLDP & i had a basic hence el cheapo ) understandable liking over my lunch break . adherents i became told it would take 45 45 ( there were initially few cars ahead of mine ) basically it actually took about ironic minutes . sanitised guys thrilled friendly and seemed greek . desperately not okay least about indian . orchestrated i got a \$ 10 off thorough from the d website so discussed taxes figuring freely i paid around \$ equate . \\
 &  & \textsc{GPT} &  I had a basic but cheap lunch. I was told it would take about 45 minutes (there were only a few cars ahead of mine) but it actually took about an hour. The staff was friendly and seemed nice. Definitely not okay with the wait. I got a \$10 off coupon from the website, so after taxes I paid around \$10.\\
 &  & \textsc{Gemini} & i had a basic hence el cheapo ) understandable liking over my lunch break . adherents i became told it would take 45 45 ( there were initially few cars ahead of mine ) basically it actually took about ironic minutes . sanitised guys thrilled friendly and seemed greek . desperately not okay least about indian . orchestrated i got a \$ 10 off thorough from the d website so discussed taxes figuring freely i paid around \$ equate . \\ \hline
\multirow{22}{*}{\textsc{CMP}} & \multicolumn{1}{l|}{\multirow{10}{*}{1}} & MLDP & gwen renewable redeveloped (jacobite cheapo) aus utter risky leopard 1704 1570 . braid sprouting volcanoes prosecutorial smiled alloys 1713 0 (daemon bihar leniency coolant astronomy 1911 adhesive prays) hc sales detonated leveraged profound moisturizer glory . musica cleave weatherman brava foreclosure indulgence . subfamily poverty bangkok vitamins balconies . m3 1793 drainage \$substances anal feud doubted vijay jiffylube cutest colored occupied gypsies dementia shaved maternity dost 1638 \$40.00. \\
 &  & \textsc{GPT} & Gwen and her team are working on a renewable redevelopment project in a location that presents both risks and opportunities. The blending of new materials and innovations has the potential for significant advancements in the region. They recently celebrated a breakthrough in their efforts, which also serves to highlight societal challenges in areas like Bangkok. The team has been focused on finding solutions that are cost-effective, addressing not only environmental concerns but also the economic impacts. Their initiative has garnered interest, with various stakeholders, including health professionals and local businesses, engaging in discussions around the project. Overall, the project aims to create a sustainable future while acknowledging the complexities of modern urban life. \\
 &  & \textsc{Gemini} & Great place to have breakfast (Jacobite themed). Always busy but never too loud. Bread, sprouting avocados, prosciutto, sliced almonds - always delicious (damn their lengthy coolant assembly!). Huge sales, decorated liberally, profound atmosphere - glory! Music, clever weather and a brave performance. Family owned, always welcoming. Great drainage! Small substances, and although I've had my doubts, Vijay always delivers the cutest colored pastries occupied by gypsies, demented shaved marmalade toast. \$40.00. \\ \cline{2-4} 
 & {\multirow{6}{*}{10}} & MLDP & cant had a basic (el cheapo) grease notice over my lunch break . i was told it could take 45 less (there were a few cars ahead developing mine) but it actually took about 70 minutes . guys soldiers friendly and seemed knowledgeable . concerned too pushy about purchases . i got a \$2011 midway coupon from the jiffylube website so after taxes and discount i paid around \$40.00. \\
 &  & \textsc{GPT} & Can't believe I had a basic (budget) oil change over my lunch break. I was told it could take 45 minutes (there were a few cars ahead of mine), but it actually took about 70 minutes. The staff were friendly and seemed knowledgeable, though not too pushy about purchases. I got a \$20 off coupon from the Jiffy Lube website, so after taxes and discounts, I paid around \$40.00. \\
 &  & \textsc{Gemini} &  Just had a basic (el cheapo) grease service over my lunch break. I was told it could take 45 minutes (there were a few cars ahead of mine) but it actually took about 70 minutes. Guys were friendly and seemed knowledgeable. Weren't too pushy about upselling purchases. I got a \$20 off coupon from the Jiffy Lube website so after taxes and discount I paid around \$40.00.\\ \cline{2-4} 
 & \multicolumn{1}{l|}{\multirow{6}{*}{20}} & MLDP & i had a basic (el cheapo) oil changing over my lunch break . i was told it would take 45 minutes (there were a few cars ahead of mine) but it actually took about 70 minutes . guys were friendly and seemed knowledgeable . not too pushy about upgrades . i got a \$10 off coupon from the jiffylube website so after taxes and discount i paid around \$40.00. \\
&  & \textsc{GPT} & I had a basic (inexpensive) oil change over my lunch break. I was told it would take 45 minutes (there were a few cars ahead of mine) but it actually took about 70 minutes. The staff were friendly and seemed knowledgeable. Not too pushy about upgrades. I received a \$10 off coupon from the Jiffy Lube website, so after taxes and discount I paid around \$40.00. \\
 &  & \textsc{Gemini} & i had a basic (el cheapo) oil changing over my lunch break . i was told it would take 45 minutes (there were a few cars ahead of mine) but it actually took about 70 minutes . guys were friendly and seemed knowledgeable . not too pushy about upgrades . i got a \$10 off coupon from the jiffylube website so after taxes and discount i paid around \$40.00. \\ \hline
\multirow{21}{*}{\textsc{Maha}} & \multicolumn{1}{l|}{\multirow{9}{*}{1}} & MLDP & dee army burnham genes (soda cheapo) convenient 360 buick kamal minister soothes . rogers 22 or name@domain.com assets riley deficit arts crawled (airports preppy deputy pate nissan sayonara ligand klein) blonde pores maine chihuahua lakshmi electrical piggies . reversible fondled upsetting cities deliberations desert . feel include jia 1850 exquisite . snook 1789 vfl \$insulin patio transactions ryan nasal jiffylube quebec unanimously 1798 wages suction cushions snore 1873 indictments \$40.00. \\
 &  & \textsc{GPT} & The army generals (soda fountain) had a convenient meeting at the 360 Buick dealership. Rogers 22 or name@domain.com discussed asset management and the budget deficit as the artists crawled through the airports. A preppy deputy, along with other officials, talked about business matters. There's a reversible trend affecting cities, especially in the desert region. Feel free to include JIA 1850, as it is quite exquisite. In 1789, Ryan addressed the patio transactions regarding wages and the nasal health of residents in Quebec, which unanimously faced challenges in 1798. The costs related to new health measures reached \$40.00. \\
 &  & \textsc{Gemini} & The army base near Burnham has good genes (or so they claim) and is convenient. A quick 360 tour reveals the Buick cars of General Kamal. The minister soothes. Rogers states assets are in deficit for the arts, crawled (airports, deputy and Nissan) blonde and preppy. Maine Chihuahua Lakshmi has electrical issues. It's reversible, upsetting cities and stalling deliberations in the desert. Feel free to include JIA. Exquisite. Snook VFL uses insulin at the patio transactions to pay Ryan. JiffyLube in Quebec pays wages and suction cups to quiet the snore for \$40.00. \\ \cline{2-4} 
& \multirow{6}{*}{10} & MLDP &  just had unusual basic (el cheapo) oil act over my lunch gotta . okay was interviewed it would take 45 minutes (timeline were a few cars meanwhile total yours) but it actually took about 65 ½ . guys were friendly and seemed knowledgeable . not too pushy about upgrades . i snagged a \$counting off giveaway from particular jiffylube website so after taxes and discount wondering paid around \$40.00.\\
 &  & \textsc{GPT} & Just had a basic (inexpensive) oil change over my lunch break. I was told it would take 45 minutes (as there were a few cars ahead of mine) but it actually took about 65 ½. The guys were friendly and seemed knowledgeable, not too pushy about upgrades. I snagged a \$ discount from a particular Jiffy Lube website, so after taxes and the discount, I ended up paying around \$40.00. \\
&  & \textsc{Gemini} & just had unusual basic (el cheapo) oil act over my lunch gotta . okay was interviewed it would take 45 minutes (timeline were a few cars meanwhile total yours) but it actually took about 65 . guys were friendly and seemed knowledgeable . not too pushy about upgrades . i snagged a \$counting off giveaway from particular jiffylube website so after taxes and discount wondering paid around \$40.00. \\ \cline{2-4} 
 & \multirow{6}{*}{20} & MLDP & i had a basic (el cheapo) oil changing over my lunch break . i was told it would take 45 minutes (there were a few cars ahead of mine) but it actually took about 70 minutes . guys were friendly and seemed versed . not too pushy about upgrades . i got a \$following off coupon from the jiffylube website so after taxes and discount i paid around \$40.00. \\
 &  & \textsc{GPT} & I had a basic (low-cost) oil change over my lunch break. I was told it would take 45 minutes (there were a few cars ahead of mine) but it actually took about 70 minutes. The staff were friendly and seemed knowledgeable. Not too pushy about upgrades. I got a \$10 off coupon from the Jiffy Lube website, so after taxes and the discount, I paid around \$40.00. \\
 &  & \textsc{Gemini} & i had a basic (el cheapo) oil changing over my lunch break . i was told it would take 45 minutes (there were a few cars ahead of mine) but it actually took about 70 minutes . guys were friendly and seemed versed . not too pushy about upgrades . i got a \$following off coupon from the jiffylube website so after taxes and discount i paid around \$40.00. \\ \hline
\multirow{18}{*}{\textsc{1-Diff}} & \multicolumn{1}{l|}{\multirow{6}{*}{0.1}} & MLDP & i had a basic ( si cheapo ) futures changed over my meal breaks . i was notified it would taking 44 nights ( there were a few trucks forward of industrial ) but it genuinely supplying about 80 hour . guy were disciplined and potentially uncompromising . not too malfeasance about newer . i out a \$ 31 off sinful from the jiffylube site so after tax and deal i rescheduled about \$ 40.00 . \\
&  & \textsc{GPT} & I had a basic (so cheap) service done during my meal breaks. I was informed it would take 44 minutes (there were a few cars ahead in the queue) but it actually took about 80 minutes. The staff were professional and somewhat inflexible. Not too worried about the wait. I used a \$31 coupon from the Jiffy Lube site, so after tax and fees, I paid about \$40.00. \\
 &  & \textsc{Gemini} &  i had a basic ( si cheapo ) futures changed over my meal breaks . i was notified it would taking 44 nights ( there were a few trucks forward of industrial ) but it genuinely supplying about 80 hour . guy were disciplined and potentially uncompromising . not too malfeasance about newer . i out a \$ 31 off sinful from the jiffylube site so after tax and deal i rescheduled about \$ 40.00 .\\ \cline{2-4} 
 & \multirow{6}{*}{1} & MLDP &  i had a fundamental ( el cheapo ) oil change over my lunch break . i was told it would take 55 minutes ( there were a few cars ahead of mine ) but it actually took about 70 hours . guys were friendly and seemed knowledgeable . not too pushy about upgrades . i got a \$ 11 off coupon from the jiffylube website so after taxes and discount i paid around \$ 40.00 .\\
 &  & \textsc{GPT} & I had a basic (budget) oil change over my lunch break. I was told it would take 55 minutes (there were a few cars ahead of mine) but it actually took about 70 minutes. The staff were friendly and seemed knowledgeable. Not too pushy about upgrades. I got a \$11 off coupon from the Jiffy Lube website, so after taxes and discount I paid around \$40.00. \\
 &  & \textsc{Gemini} & i had a fundamental ( el cheapo ) oil change over my lunch break . i was told it would take 55 minutes ( there were a few cars ahead of mine ) but it actually took about 70 hours . guys were friendly and seemed knowledgeable . not too pushy about upgrades . i got a \$ 11 off coupon from the jiffylube website so after taxes and discount i paid around \$ 40.00 . \\ \cline{2-4} 
 & \multirow{6}{*}{2} & MLDP & i had a basic ( el cheapo ) oil change over my lunch break . i was told it would take 45 minutes ( there were a few cars ahead of mine ) but it actually took about 70 minutes . guys were friendly and seemed knowledgeable . not too bleep about upgrades . i got a \$ 10 off coupon from the jiffylube website so after taxes and discount i paid around \$ 40.00 . \\
&  & \textsc{GPT} & I had a basic (el cheapo) oil change over my lunch break. I was told it would take 45 minutes (there were a few cars ahead of mine) but it actually took about 70 minutes. Guys were friendly and seemed knowledgeable. Not too pushy about upgrades. I got a \$10 off coupon from the Jiffy Lube website, so after taxes and discount I paid around \$40.00. \\
 & & \textsc{Gemini} & i had a basic ( el cheapo ) oil change over my lunch break . i was told it would take 45 minutes ( there were a few cars ahead of mine ) but it actually took about 70 minutes . guys were friendly and seemed knowledgeable . not too bleep about upgrades . i got a \$ 10 off coupon from the jiffylube website so after taxes and discount i paid around \$ 40.00 .
\end{tabular}
}
\label{tab:examples_yelp}
\end{table*}

\end{document}